\documentclass[aps,prl,reprint]{revtex4-2}

\usepackage{amsmath}
\usepackage{amssymb}

\usepackage{xcolor}

\usepackage{mathtools}
\usepackage{stmaryrd}
\newtheorem{theorem}{Theorem}
\newtheorem{definition}{Definition}

\newtheorem{corollary}{Corollary}[theorem]

\newcommand{\x}{\boldsymbol{x}}
\newcommand{\pa}{\mathrm{pa}}
\newcommand{\nin}{\mathrm{nin}}
\newcommand{\I}{\mathrm{I}}
\newcommand{\ch}{\mathrm{ch}}

\newcommand{\qqt}{q'}
\newcommand{\xx}{X}

\newcommand{\MMC}{\mathcal{MC}}
\newcommand{\N}{\mathbb{N}}
\newcommand{\R}{\mathrm{R}}

\newcommand{\size}[1]{\mathbf{SIZE}(#1)}

\newcommand{\MMClass}[1]{\mathbf{MMC}(#1)}
\newcommand{\C}{\mathcal{C}~}
\newcommand{\ADD}{\text{ADD}}
\newcommand{\Poly}{\mathbf{P\text{/poly}}}
\newcommand{\NC}{\mathbf{NC}}
\newcommand{\poly}{\mathrm{poly}}

\newcommand{\basisg}{\Tilde{\mathcal{B}}}
\newcommand{\basis}{\mathcal{B}}

\newcommand{\inn}{\mathrm{in}}
\usepackage[colorlinks=true,allcolors=blue]{hyperref}

\begin{document}
\title{A strictly positive lower bound on the thermodynamic cost of running a Boolean circuit}

%Title suggestion
%\title{Mismatch cost as a thermodynamic resource for Boolean circuits}

\author{Abhishek Yadav${}^{1, 2, \dagger}$}
\author{Mahran Yousef${}^{3, \dagger}$}
\thanks{Corresponding author: mahran.yousef2@gmail.com}
\author{David Wolpert${}^{1, 4, 5, 6}$}

\affiliation{\vspace{.25cm}${}^1$Santa Fe Institute, 1399 Hyde Park Road
Santa Fe, NM 87501, USA}
\affiliation{${}^2$Department of Physical Sciences, IISER Kolkata, Mohanpur 741246, India}
\affiliation{${}^3$Independent Researcher}
\affiliation{${}^4$Complexity Science Hub, Vienna, Austria}
\affiliation{${}^5$Arizona State University, Tempe, AZ 85281, USA}
\affiliation{${}^6$International Center for Theoretical Physics, Trieste 34151, Italy}
\affiliation{${}^\dagger$Equal Contributions}
%\affiliation{${}^*$Corresponding author: mahran.yousef2@gmail.com}
\begin{abstract}
    All digital computers implement input-output functions using logic gates connected into circuits. Different circuits computing the same function may nevertheless incur different resource costs, and circuit complexity theory studies these costs through measures such as the number of gates and the length of the longest path from input to output. Energetic cost is another important resource, however, that is typically not included among these measures. To address this, we use mismatch cost (MMC): a nonnegative contribution to the entropy production of a process that can be characterized largely independently of the detailed physical implementation of that process, providing a natural way to analyze the thermodynamic cost of Boolean circuits at an abstract, computational level. We derive an expression for the MMC of circuits composed of Boolean gates, relate it to standard complexity measures such as circuit size and depth, and use it to define \emph{mismatch cost complexity} as a measure of thermodynamic resource cost. For Boolean circuits computing a non-constant Boolean function, this expression also implies a strictly positive MMC for every input distribution, and therefore a strictly positive lower bound on total entropy production. We characterize when MMC scales linearly with circuit size and when it does not, and compare the MMC of different circuit families that compute the same Boolean function. Together, these results lay the foundation for treating mismatch cost as a resource within circuit complexity theory.
\end{abstract}
% \begin{abstract}
% All digital computers are based on Boolean circuits doing mathematical operations on binary inputs. The field of circuit complexity theory focuses on the resource costs such as circuit's size 
% (its number of gates) and depth 
% (the longest path length from the circuit's input to its output)

% Traditionally,
% circuit complexity theory has focused on 
% the resource costs of a circuit's size 
% (its number of gates) and depth 
% (the longest path length from the circuit's input to its output). {\color{blue}
% In this paper, we extend circuit complexity theory by studying the mismatch cost (MMC) of a circuit that is run repeatedly to quantify the part of its EP caused
% by a mismatch between the actual distribution over circuit states and the prior distribution
% for which the corresponding physical process is least dissipative. MMC is not, in general,
% the complete EP of a physical device; rather, it is a nonnegative contribution to EP and
% therefore a rigorous lower bound on the device's thermodynamic cost. We discuss conditions
% under which the MMC of a circuit is proportional to circuit size and conditions under which
% that proportionality fails. We also compare the MMCs of different circuit families that
% implement the same family of Boolean functions, and analyze how differences among the priors
% associated with gate types affect the circuit-level lower bound. These results lay the
% foundation for extending circuit complexity theory to include mismatch cost as a resource
% cost.
% }

% \end{abstract}

\maketitle

\section{Introduction}

\subsection{Background and scope}

In his foundational work~\cite{turing1936computable}, Alan Turing introduced the concept of Turing machines and demonstrated that functions can be classified as either computable or non-computable. Within the class of computable functions, it was later shown that some inherently require more computational time than others~\cite{cook2007overview, hartmanis1965computational}. The study of what it means for one function to be more difficult to compute than another is the subject of computational complexity theory~\cite{arora2009computational}, which provides complexity measures in terms of resources such as time and memory within a given model of computation.

Boolean circuits provide one such model of computation. They characterize the complexity of a function in terms of the size and depth of the circuit that implements it: the size of a circuit is its total number of gates, while its depth is the length of the longest path from an input to an output, representing the time required for computation~\cite{arora2009computational, vollmer1999introduction}. A given Boolean function can be implemented by infinitely many circuits. Identifying the most efficient one among them is both theoretically difficult and practically important, and circuit complexity theory provides a framework for comparing and optimizing circuits on this basis.

Energetic cost, though a natural candidate for such a resource, has not been as thoroughly integrated into complexity theory. Early discussions of complexity measures did consider physical energy expenditure and thermodynamic work as a potential metric~\cite{cobham1965intrinsic}. Landauer and Bennett later used equilibrium statistical mechanics to argue that erasing a single bit of information necessarily generates at least $k_B T \ln 2$ of heat, where $k_B$ is Boltzmann's constant and $T$ is the temperature of the surrounding heat bath~\cite{landauer1961irreversibility, bennett1982thermodynamics}. However, computation is inherently a non-equilibrium process, and modeling it within equilibrium thermodynamics left this early analysis semi-formal and limited in scope. The theoretical tools needed to rigorously analyze systems driven far from equilibrium were not yet available, and the thermodynamics of computation remained an open and underdeveloped area for decades.

More recently, advances in non-equilibrium thermodynamics, particularly within the framework of stochastic thermodynamics, have made it possible to rigorously define and analyze thermodynamic quantities such as work, entropy, and heat dissipation in systems driven far from equilibrium~\cite{van2015ensemble, seifert2012stochastic, seifert2018stochastic, esposito2010entropy, esposito2012stochastic}. These tools are now being applied to fundamental questions about the thermodynamic costs of computation and communication processes~\cite{van2015ensemble, seifert2012stochastic, wolpert2019stochastic, wolpert2024thermodynamics, wolpert2024stochastic}. %{\color{black}In particular, stochastic thermodynamics allows us to analyze entropy production, a nonnegative measure of thermodynamic irreversibility and dissipation of thermodynamic work.}

One of the central quantities in stochastic thermodynamics is entropy production, a nonnegative quantity that measures the thermodynamic irreversibility of a process and is closely related to the associated dissipated work~\cite{seifert2012stochastic}. Beyond providing a generalization of the second law of thermodynamics to systems driven far from equilibrium, a large body of work in stochastic thermodynamics has sought to identify and bound contributions to entropy production arising from constraints on the underlying physical dynamics. Examples include thermodynamic uncertainty relations~\cite{barato2015thermodynamic,gingrich2016dissipation} and thermodynamic speed limits~\cite{shiraishi2018speed}. Such results establish powerful connections between dissipation and properties such as fluctuations, currents, and the time required to carry out a process, and have consequently played an important role in the study of the thermodynamic costs of information processing~\cite{wolpert2024stochastic}. However, they typically require specifying substantial structure of the underlying dynamics, often through assumptions such as Markovianity, weak coupling, or detailed balance. This can make them difficult to apply when a system is described only at a more abstract, computational level.
Mismatch cost provides a complementary decomposition of entropy production. Consider a physical system with discrete state space $\mathcal{X}$ undergoing stochastic evolution over some fixed interval of time. Let $G(x_1 \mid x_0)$ denote the conditional probability that the system ends the process in state $x_1 \in \mathcal{X}$, given that it started in state $x_0 \in \mathcal{X}$. Given an initial distribution $p_X$ over $\mathcal{X}$, the system's state at the end of the process is distributed according to $p'_X(x_1) = \sum_{x_0 \in \mathcal{X}} G(x_1 \mid x_0)\, p_X(x_0)$, which we write as $Gp_X$ for short. The total EP associated with an initial distribution $p_X$, denoted as $\sigma(p_X)$, can always be expressed as~\cite{kolchinsky2017dependence, kolchinsky2021dependence}, 
\begin{align}\label{eq:single_island_mismatch1}
\sigma(p_X) = \MMC(p_X) + \sigma(q_X),
\end{align}
where 
\begin{equation}\label{eq:single_island_mismatch2}
    \MMC(p_X) = D(p_X|q_X) - D(Gp_X|Gq_X),
\end{equation}
and $q_X$ is an initial distribution that minimizes the EP of the process. Thus, $\sigma(q_X)$ is called the \emph{residual EP} and it is the minimum EP achievable by that process. $\MMC(p_X)$ is called the \emph{mismatch cost} and it is the additional EP incurred when the process is instead run on the distribution $p_X$. By the data-processing inequality, $\MMC(p_X)\geq 0$.

An important feature of this decomposition is that the mismatch cost depends only on the input distribution $p_X$, the EP-minimizing distribution $q_X$, and the stochastic map $G$, rather than on the detailed dynamical trajectory by which $G$ is physically implemented. Consequently, once $q_X$ is specified, MMC can be studied directly at the level of the computation itself. This makes it particularly useful for analyzing the thermodynamic costs of information-processing systems for which the logical transformation is known but the microscopic physical realization is either unspecified or deliberately abstracted away. Mismatch cost has accordingly been applied to settings including communication channels~\cite{yadav2025minimal}, stored-program computers~\cite{yadav2026entropy}, and other information-processing systems~\cite{manzano2024thermodynamics}. In the present work, we exploit this implementation-independent character of mismatch cost to study Boolean circuits. Rather than committing to a particular physical realization of each logical gate, we ask how much mismatch cost is induced by the computational structure of the circuit itself, and how this cost relates to familiar circuit-complexity measures such as size and depth.

This perspective is useful for both complexity theory and circuit design. From a complexity-theoretic viewpoint, it provides a way to place a thermodynamic resource alongside familiar resources such as circuit size and depth. From an engineering viewpoint, the design of digital circuits already involves trade-offs among resources such as area, delay, and energy consumption~\cite{horowitz2014computing,weste2015cmos}, while energy use and the associated heat dissipation are increasingly important constraints~\cite{theis2017end,masanet2020recalibrating}. Existing circuit-level energy analyses are typically tied to particular device technologies, relying, for example, on CMOS power models or electrical dissipation in interconnects~\cite{weste2015cmos,chandrakasan1995minimizing,bellaouar2012low}. Mismatch cost provides a complementary perspective: it identifies a thermodynamically grounded contribution to dissipation that can be analyzed without specifying such microscopic hardware details. This makes it possible to ask how properties of the computation itself---including circuit topology, gate composition, modularity, and degree of parallelism---constrain this contribution to thermodynamic cost.

\subsection{Summary of our contributions and roadmap}

Our starting point is that a Boolean circuit diagram specifies the logical dependencies among gates, but does not by itself specify the dynamics of the joint physical state of the circuit. We therefore introduce a simple execution model in which the same circuit is used repeatedly. At the beginning of each run, the input nodes are re-initialized with a fresh sample from an input distribution $p_{\inn}$, while all non-input gates retain the states produced during the preceding run. The non-input gates are then updated according to a topological ordering of the circuit, either one gate at a time or, more generally, one layer at a time. A gate changes state only when it is executed; until then, its state is preserved from the previous run. Consequently, during an execution the circuit passes through a sequence of joint distributions in which some gates have already been updated using the current input while the remaining gates still carry states inherited from the previous computation.

For this execution model, we derive an expression for the total mismatch cost incurred during a complete run of the circuit. The cost includes both the re-initialization of the inputs and the subsequent execution of the non-input gates. Although the mismatch cost associated with an individual gate update initially involves global properties of the circuit state, these global entropy terms telescope when summed over a complete run. The resulting expression decomposes the total mismatch cost into terms determined by the input distribution, the priors associated with the input and gate-update processes, and the statistical dependencies generated locally between gates and their parents. We also derive the corresponding expression for layer-by-layer execution. Since layer-by-layer execution is a temporal coarse-graining of gate-by-gate execution, its mismatch cost is no greater than that of the corresponding gate-by-gate implementation.

We then use this circuit-level expression to introduce \emph{mismatch cost complexity}, which treats MMC as an additional resource measure for circuit families, alongside conventional measures such as size and depth. The definition is made relative to a circuit basis together with the priors associated with its gate types, and measures the worst-case mismatch cost over input distributions. We establish a general upper bound on this cost in terms of the number of gates and their local priors. In particular, when the relevant conditional prior probabilities are uniformly bounded away from zero, the worst-case MMC is bounded above by a quantity linear in the circuit size, up to the contribution from re-initializing the inputs. As a consequence, under the stated assumptions, every function family computable by polynomial-size circuits also admits polynomial mismatch cost, giving
\[
\mathrm{P/poly}\subseteq \mathrm{MMC(poly)}.
\]
We further extend the bound to heterogeneous circuits in which different gate types have different priors, showing that the mismatch cost can depend on the composition of gate types and not merely on the total number of gates.

These results also show why MMC constitutes a resource distinct from conventional circuit size or depth. Circuits of different architectures that implement the same Boolean function can incur different mismatch costs under the same input distribution and prior basis. We illustrate this by comparing ripple-carry and carry look-ahead implementations of binary addition, and by examining how heterogeneous gate priors can alter the relationship between circuit size and MMC. More generally, minimizing circuit size need not minimize mismatch cost: a larger circuit composed of gate types whose priors are better matched to the distributions encountered during computation can have lower MMC than a smaller implementation of the same function.

The remainder of the paper is organized as follows. In Sec.~\ref{sec2}, we review the necessary background on stochastic thermodynamics, mismatch cost, and circuit complexity theory. In Sec.~\ref{sec:3}, we specify the dynamics of repeated circuit execution and derive the total MMC for both gate-by-gate and layer-by-layer implementations. In Sec.~\ref{sec:4}, we introduce mismatch cost complexity and establish general bounds relating it to circuit size, including extensions to heterogeneous gate priors. In Sec.~\ref{sec:5}, we examine the implications of these results for comparing and optimizing circuit families that implement the same Boolean function. Finally, in Sec.~\ref{sec:6}, we discuss the interpretation and limitations of MMC as an implementation-independent contribution to entropy production, its relationship to logically reversible computation, and several directions for future work.

\subsection{Earlier work}

The tools of stochastic thermodynamics have been applied to computational models such as finite automata~\cite{wolpert2024thermodynamics, kardecs2022inclusive, manzano2024thermodynamics} and Turing machines~\cite{strasberg2015thermodynamics, kolchinsky2020thermodynamic}, providing one of the first principled frameworks for analyzing the thermodynamic costs of computation. Specific to circuits, prior work has estimated the heat dissipated during gate operations using Landauer’s bound~\cite{Gershenfeld_1996}. While this was a pioneering approach, stochastic thermodynamics now recognizes that the Landauer bound—which accounts only for entropy changes—can be zero even when the total dissipated heat is substantial~\cite{sagawa2014thermodynamic}. 

More recent studies have incorporated EP into the analysis of Boolean circuits by modeling gate dynamics as continuous-time Markov chains (CTMCs)~\cite{wolpert2020thermodynamics}. However, such a framework requires properly incorporating dynamics over hidden states, which in general are necessary for there to be a CTMC that can implement discrete-time dynamics~\cite{wolpert2019space}. In contrast, our analysis does not assume any specific form for the underlying continuous-time dynamics. Furthermore, in our model, the logical values at gates are not reset between successive runs, more closely reflecting the behavior of many physical computing devices. In contrast, prior work~\cite{wolpert2020thermodynamics} assumes a full reset after each computation.

Another closely related set of results analyzes entropy production in stochastic processes structured as Bayesian networks~\cite{ito2013information, ito2016information, wolpert2020uncertainty}. These works provide general decompositions of entropy production in terms of information flows along the edges of a directed acyclic graph. The formalism developed in those works is in principle extensible to circuit-like architectures, and our framework can be viewed as a specialization of these more general thermodynamic decompositions to Boolean circuits, together with a complexity-theoretic perspective on how costs scale with circuit size.

% {\color{blue}
% This paper presents several contributions concerning the mismatch component of the
% thermodynamic cost of Boolean circuits. First, we derive a circuit-level MMC and therefore a
% lower bound on the EP generated when a circuit is run. This bound arises from the modular
% structure of circuits: each update acts on only a subset of the circuit while the full circuit
% state generally contains correlations extending beyond that subset. We then introduce
% \emph{mismatch cost complexity}, establish formal relationships between MMC complexity and
% traditional size complexity, and compute MMC for several circuit families implementing the
% same Boolean function. These comparisons concern the lower-bound contribution represented
% by MMC, not the unknown total EP of every possible hardware realization.

% The resulting complexity classes are defined relative to a specified \emph{basis with
% priors}. The allowed gate basis and its associated priors together specify the thermodynamic circuit model,
% just as the allowed gate basis specifies the logical circuit model in ordinary circuit
% complexity. Accordingly, our classes organize functions by the scaling of worst-case MMC for
% that specified prior basis; they are not claimed to be universal complexity classes for total
% entropy production across all physical implementations.
% }

\section{Preliminaries}\label{sec2}

\subsection{Stochastic Thermodynamics and Mismatch Cost}\label{sec:2b}

Consider a system with a discrete state space evolving stochastically over a fixed time interval, with the resulting initial state to final state map specified by a conditional distribution $G$. For an initial distribution $p_X$, the total entropy production associated with the process can be written as~\cite{van2015ensemble,kolchinsky2017dependence},
\begin{equation}\label{EP_def}
    \sigma(p_{X}) = Q(p_X) - [S(Gp_X) - S(p_X)],
\end{equation}
where $Q(p_X)$ is called the \textit{total entropy flow} and it corresponds to the change in the thermodynamic entropy of the environments during the process. If the system interacts with $N$ heat reservoirs, the total entropy flow is given by the sum of the energy exchanged with each reservoir, weighted by the inverse of its temperature:
\begin{equation}
    Q = \sum_{i=1}^N \frac{\mathcal{Q}_i}{k_B T_i},
\end{equation}
where $\mathcal{Q}_i$ represents the net energy flow into the $i$-th reservoir at temperature $T_i$, and $k_B$ is Boltzmann’s constant.

Central to the formulation of mismatch cost is the notion of \emph{islands} associated
with a given computational map $G$. Define a relation on the state space $\mathcal{X}$ by
\begin{equation}
    x \sim x' \iff \exists\, y \in \mathcal{X} : G(y\mid x) > 0 \text{ and }
    G(y \mid x') > 0,
\end{equation}
and let $\mathcal{L}_G(\mathcal{X})$ denote the partition of $\mathcal{X}$ induced by the
transitive closure of this relation. We refer to $\mathcal{L}_G(\mathcal{X})$ as the
\emph{island decomposition} of $\mathcal{X}$ under $G$, and to each subspace $c \in
\mathcal{L}_G(\mathcal{X})$ as an \emph{island}. 

Within any given island $c$, define
\begin{equation}
    q_{X|c} \in \arg\min_{r\,:\,\mathrm{supp}(r) \subseteq c} \sigma(r),
\end{equation}
the distribution restricted to island $c$ that minimizes the EP in
Eq.~\eqref{EP_def}. It is shown in the supplementary information of~\cite{yadav2026entropy} that $q_{X|c}$ is unique and has full support over the island $c$. We refer to $q_{X|c}$ as the \emph{prior} distribution
associated with island $c$.

Given the set of island priors $\{q_{X|c}\}_{c \in \mathcal{L}_G(\mathcal{X})}$, and for an
arbitrary initial distribution $p_X \in \Delta_\mathcal{X}$, the EP generated by running $G$ on
$p_X$ admits the exact decomposition
\begin{equation}\label{eq:general_mismatch}
    \sigma(p_X) = \MMC(p_X) + \sum_{c \in \mathcal{L}_G} p(c)\, \sigma(q_{X\mid c}),
\end{equation}
where $p(c) := \sum_{x \in c} p_X(x)$ denotes the total probability $p_X$ assigns to island $c$,
and
\begin{align}
    \MMC(p_X) &:= D(p_X || q_X) - D(Gp_X || Gq_X) \\
    = \sum_{c \in \mathcal{L}_G}& p(c) \left[D(p_{X | c} || q_{X | c}) -
    D(Gp_{X | c} || Gq_{X | c})\right],
\end{align}
where $p_{X | c}$ is the conditional distribution of $p_X$ restricted to island $c$, and
$q_X$ is \emph{any} distribution obtained by mixing the island priors $q_{X | c}$ with
arbitrary positive weights across islands, $q_X(x) = \sum_c q(c)\, q_{X | c}(x)$ for some choice of
$\{q(c)\}_{c \in \mathcal{L}_G(\mathcal{X})}$. As the second line above makes clear, the value
of $\MMC(p_X)$ does not depend on this choice of weights $\{q(c)\}$: each term in the sum is
fully determined by $p(c)$ and the island priors $q_{X | c}$ alone, so any choice of mixing
weights for $q_X$ yields the same $\MMC(p_X)$. 

When $G$ is deterministic, its islands are simply the nonempty preimages $G^{-1}(y)$ of its outputs $y$. Consequently, for a deterministic Boolean gate, whose execution updates only the gate output while leaving its parent states fixed, each island corresponds to a fixed parent configuration, as we discuss in more detail in Sec.~\ref{sec:3c}. 

In the special case where $G$ induces a single island, i.e., $\mathcal{L}_G(\mathcal{X}) =
\{\mathcal{X}\}$, Eq.~\eqref{eq:general_mismatch} reduces to the following simpler form:
\begin{align}\label{eq:single_island_mismatch}
    \sigma(p_X) &=  \MMC(p_X) + \sigma(q_X), \\
    \MMC(p_X) &= D(p_X\|q_X) - D(Gp_X\|Gq_X),
\end{align}
where $q_X$ is now a single, unique, full-support distribution over all of $\mathcal{X}$, and
$\MMC(p_X)$ is the excess EP incurred by running the system on $p_X$ rather
than on the globally optimal $q_X$. 

The residual term $\sigma(q_X)$ depends on the specific microscopic dynamics used to implement $G$, and can vary substantially across different physical realizations of the same computation. Different realizations of the same logical process can therefore have the same mismatch cost while differing considerably in their total EP. Throughout this paper, the MMC should accordingly be understood as a nonnegative, implementation-independent lower bound on EP -- not as a claim about the total energetic cost of any specific device.

Moreover, we emphasize that thermodynamic irreversibility, as quantified by entropy production, is distinct from logical irreversibility. As shown in~\cite{sagawa2014thermodynamic}, logical and thermodynamic reversibility are independent notions. In particular, logical irreversibility---or logical reversibility---does not by itself determine the entropy production of a physical process. We return to this distinction, including its relation to fully reversible computation, in Sec.~\ref{sec:6}.

%{\color{red} Add a paragraph including broader discussion of MMC.}
\begin{figure}
    \centering
    \includegraphics[width=0.7\linewidth]{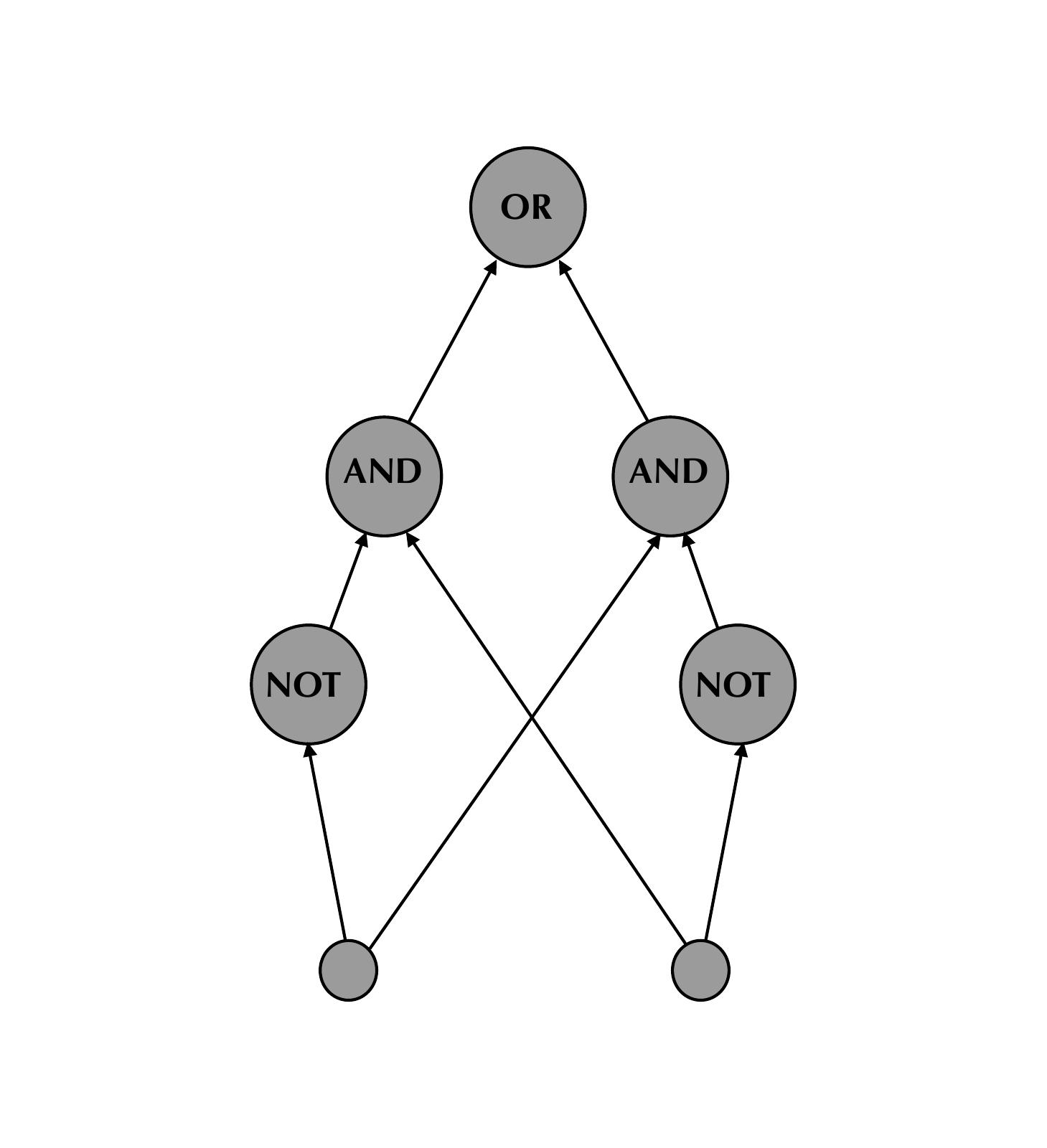}
    \caption{An example of a 2-input and 1-output circuit with a topological ordering. %The [] are input nodes and [] is the output node. 
    }
    \label{example_circuit}
\end{figure}

\subsection{Circuit complexity theory}\label{sec:2a}

In this section, we provide a brief and necessarily compact overview of basic concepts from circuit complexity theory. For a more detailed and comprehensive treatment, see~\cite{arora2009computational}. 

A circuit is built from basic logic gates that are interconnected to process inputs and produce outputs (Fig.~\ref{example_circuit}). A \textit{basis} is a set $\mathcal{B}$ of Boolean functions that define the allowable gates in a circuit. For example, a basis containing AND and NOT gates is considered a \textit{universal basis} because any Boolean function can be implemented using only these gates. Formally, a Boolean circuit is represented as a \textit{directed acyclic graph} (DAG), denoted by $(V, E, \mathcal{B}, \mathcal{X})$, where $V$ is the set of nodes representing logic gates selected from a finite basis $\mathcal{B}$, $E$ is the set of edges connecting the gates, and $\mathcal{X}$ is a joint state space $\bigotimes_{g \in V} \mathcal{X}_{g}$ with each $\mathcal{X}_{g}$ representing the set of possible states associated with gate $g$. 

The direction of edges reflects dependencies between gates. A node $g$ has incoming edges from nodes known as its \textit{parent nodes}, denoted as $\pa(g)$, and outgoing edges to nodes known as its \textit{children nodes}, denoted as $\ch(g)$. Nodes that have no incoming edges, called \textit{input nodes} or \textit{root nodes}, serve as the circuit's inputs. We denote the set of input nodes as $V_{\inn} \subset V$ and the set of all non-input nodes as $V_{\nin} = V \setminus V_{\inn}$. The output state of each gate, $\x_g \in \mathcal{X}_g$, depends on the states of its inputs,
given by its parents' output states $\x_{\pa(g)} \in \mathcal{X}_{\pa(g)}$. This dependence is
captured by a conditional distribution $\pi_g(\x_g \mid \x_{\pa(g)})$ associated with each gate
$g$. A gate is deterministic if, for every parent state $\x_{\pa(g)}$, $\pi_g(\cdot \mid
\x_{\pa(g)})$ places all of its probability on a single output state; otherwise, the gate is
stochastic, and $\pi_g(\cdot \mid \x_{\pa(g)})$ may assign positive probability to
multiple output states for a given parent state.

% \color{red} The thermodynamic derivations below allow such stochastic gate kernels. Unless explicitly stated otherwise, however, the circuit-complexity statements about exact computation of Boolean functions, including the definitions of the associated complexity classes, are restricted to deterministic logical gates.

The \textit{size} of a circuit is the total number of gates in it. The \textit{depth} of a circuit is defined as the length of the longest path from an input node to an output node. Circuit size and depth are two important complexity measures of a circuit. They also provide a measure of the hardness of computation of a function $f$ as the size and depth of the minimal circuit that computes it. 

An $n$-input Boolean function is a mapping  
\[
f^n: \{0,1\}^n \to \{0,1\}^m,
\]  
which takes $n$-bit binary inputs and produces $m$-bit binary outputs. A \textit{family of Boolean functions} is defined as a sequence $f = \{f^n\}_{n\in \mathbb{N}}$, where each $f^n$ corresponds to a specific input size $n$. This formulation allows us to define an associated function  
\[
f: \{0,1\}^* \to \{0,1\}^*,
\]  
that operates on inputs of arbitrary length. For example, consider the addition function  
\[
\ADD^n: \{0,1\}^{2n} \to \{0,1\}^{n+1}
\]  
which takes two $n$-bit binary representations of natural numbers and outputs their sum in binary. The function family $\{\ADD^n\}_{n \in \mathbb{N}}$ extends this addition operation to all possible input sizes.

Circuits differ fundamentally from other models of computation. In models like Turing machines, a single machine accepts input of any length. In contrast, circuits require a distinct specification for each input length. This means that to compute functions belonging to the same family but with different input sizes, one must design distinct circuits for each case. Because the model does not provide a uniform mechanism for handling all input lengths, it is classified as a \textit{non-uniform} model of computation~\cite{arora2009computational, vollmer1999introduction}. As a result, computation in this model is described using a \textit{circuit family}, denoted as $\{C_n\}_{n\in \mathbb{N}} $, where each circuit $C_n$ corresponds to inputs of length $n$. A circuit family $\C$ is a sequence $\{C_n\}_{n \in \N}$ of Boolean circuits, where $C_n$ is the circuit with $n$ inputs.

\noindent
A circuit family $\mathcal{C}$ is said to compute a function $f$ if, for every input string $w$ of length $n$, the circuit correctly evaluates the function:  
\[
    f^n(w) = C_n(w) 
\]
for all $n \in \mathbb{N}$.  
The \textit{size complexity} of a circuit family is the number of gates in $C_n$ as a function of $n$, while the \textit{depth complexity} is the length of the longest directed path from an input node to an output node, also expressed as a function of $n$.

\begin{definition} [\textbf{Size and depth complexity}]
    Given a function $s : \N \to \N$, we say $\{C_n\}_{n \in \N}$ has \emph{size complexity} $s(n)$
    if $|C_n| = O(s(n))$, i.e.\ there exist constants $c > 0$ and $n_0 \in \N$ such that
    $|C_n| \le c \cdot s(n)$ for all $n \ge n_0$.
    
    Analogously, given a function $d : \N \to \N$, we say $\{C_n\}_{n \in \N}$ has \emph{depth
    complexity} $d(n)$ if $\mathrm{depth}(C_n) = O(d(n))$.
\end{definition}

Likewise, one sometimes speaks of the size and depth complexity of a function $f$, meaning the size and depth of the smallest circuit family (within the chosen circuit model) that computes $f$.

\begin{definition}[\textbf{Circuit complexity class}]
    Let $s : \N \to \N$ and let $f : \{0,1\}^* \to \{0,1\}$. We say $\mathrm{SIZE}(s)$ is the class
    of functions $f$ for which there exists a circuit family $\mathcal{C} = \{C_n\}_{n \in \N}$ of
    size complexity $s$ such that $\mathcal{C}$ \emph{computes} $f$, i.e.\ $C_n(x) = f(x)$ for every
    $n \in \N$ and every $x \in \{0,1\}^n$.
    
    Analogously, let $d : \N \to \N$. We say $\mathrm{DEPTH}(d)$ is the class of functions $f$ for
    which there exists a circuit family $\mathcal{C} = \{C_n\}_{n \in \N}$ of depth complexity $d$
    such that $\mathcal{C}$ computes $f$.
\end{definition}

The growth of a circuit’s size or depth with respect to the input length of a function provides a crucial measure of the function’s computational complexity. %In his seminal work, Shannon proved that most Boolean functions require circuits of exponential size, specifically showing that there exist functions that cannot be computed by circuits smaller than $2^n / 10n$. In other words, the vast majority of functions—often referred to as ``hard'' functions—such that $f \notin \size{2^n / 10n}$ and thus require exponentially large circuits to compute~\cite{shannon1949synthesis}. However, despite this result, explicitly identifying such a function has remained challenging, as most functions of practical interest can still be computed by circuits of reasonable, albeit large, size.  
An interesting class of functions emerges when restricted to ``small'' circuits, i.e., circuit families whose size does not grow faster than polynomial with input length. $\Poly$ is the class
of functions for which there exists a polynomial-size circuit family,
\begin{equation*}
    \Poly := \bigcup_{c \ge 0} \size{n^c}.
\end{equation*}
The standard class $\NC^d$ consists of functions that can be computed by a suitably uniform circuit family of polynomial
size, bounded fan-in (typically fan-in $2$), and depth $O(\log^d n)$, for some $d \in \N$. Since
the depth of a circuit corresponds to its parallel running time, and this grows only as
$(\log n)^d$, functions in $\NC^d$ can be computed very quickly in parallel, using a polynomial
amount of hardware. These classes form the $\NC$ hierarchy,
\[
\NC := \bigcup_{d \ge 1} \NC^d,
\]
which captures efficiently parallel computation under the usual uniformity requirement~\cite{arora2009computational, vollmer1999introduction}. While this hierarchy characterizes computation purely in terms of size and depth, it says nothing about the thermodynamic cost of realizing these circuits. To reason about this cost, we turn to the framework of stochastic thermodynamics.

\begin{figure*}
    \centering
    \includegraphics[trim={0 20cm 0 0}, width=1\textwidth]{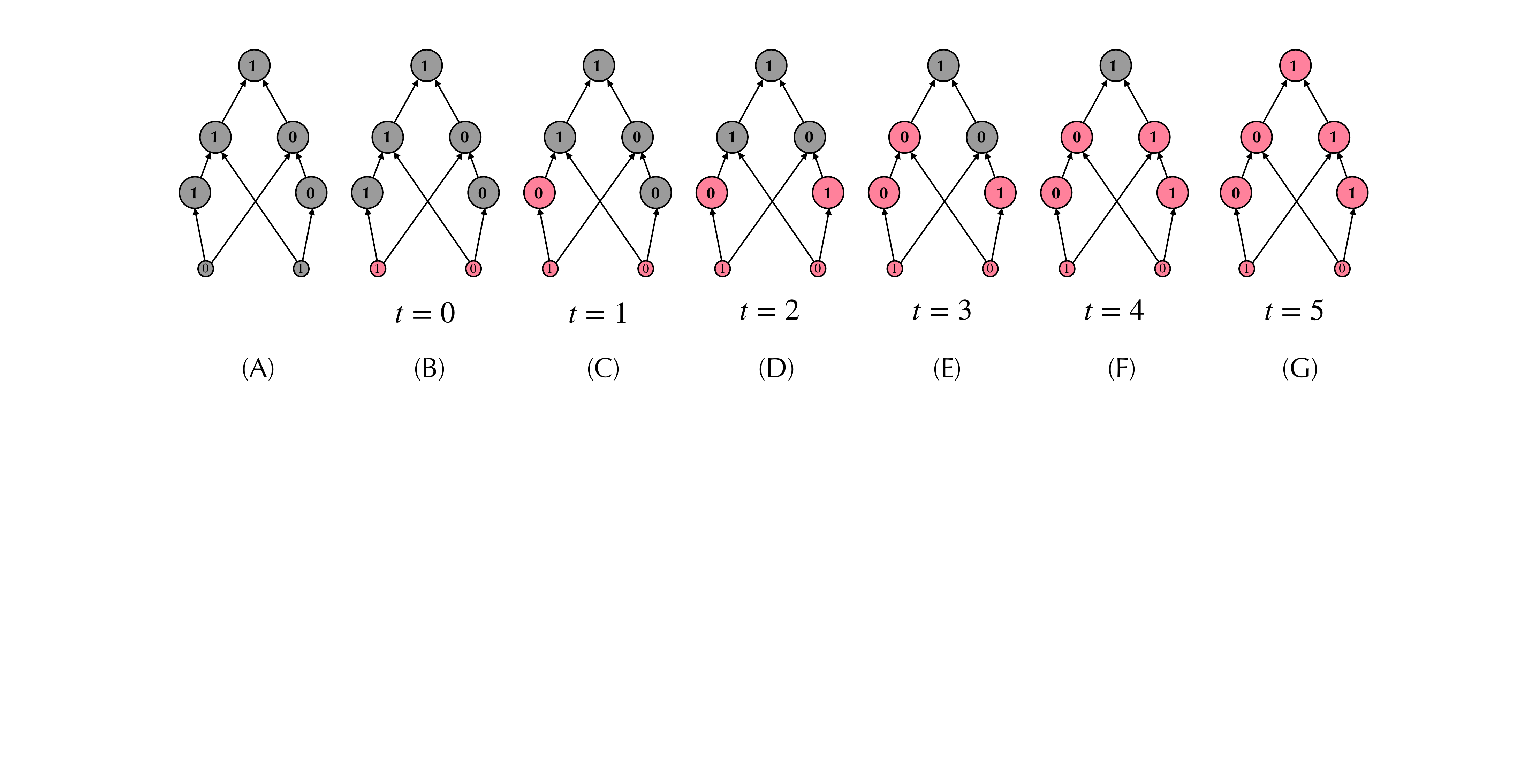}
    \caption{State space dynamics of a circuit computation.
(A) The circuit begins in a state inherited from the previous run, with all gate values logically dependent according to the circuit structure.  
(B) The inputs are updated for the new run, resetting their values independently of the rest of the circuit. 
(C–G) Gates are sequentially updated based on their input dependencies, progressively building logical correlations as computation unfolds.}
\label{circuit_dynamics}
\end{figure*}

\section{Mismatch cost of computing with circuits}\label{sec:3}

\subsection{Dynamics of the circuit}\label{sec3a}

A circuit diagram specifies the dependency structure among its gates, but by itself does not specify the dynamics of the joint state of the physical circuit. In particular, it does not specify the order in which gates are executed or how the state of the circuit changes during an execution. In this section, we introduce a simple dynamical specification that will be used throughout the paper.

Given a circuit, choose an ordering of the non-input gates, $g_1,g_2,\ldots,g_{|V_{\nin}|}$, such that whenever there is an edge from $g_i$ to $g_j$, we have $i<j$. Such an ordering is called a \emph{topological ordering}. A run of the circuit proceeds in two stages. First, at execution step $t=0$, the input nodes are re-initialized with values drawn from the input distribution $p_{\inn}$. The non-input gates are then executed sequentially according to the chosen topological ordering: at execution step $t=1,\ldots,|V_{\nin}|$, gate $g_{t}$ is executed. At any intermediate stage of a run, gates that have already been executed have their states generated during the current run, while gates that have not yet been executed still retain their states from the previous run. 

The update of gate $g$ is specified by a conditional distribution $\pi_g(\x_g\mid \x_{\pa(g)}),$ which gives the distribution of its new output state conditioned on the current output states of its parent gates. Together, these conditional distributions determine the distribution of all non-input gates conditioned on the circuit inputs,
\begin{equation}
\R(\x_{\nin}\mid\x_{\inn})
=
\prod_{g\in V_{\nin}}
\pi_g(\x_g\mid\x_{\pa(g)}).
\end{equation}

The circuit is run repeatedly: after finishing a complete run, it is reused on new input values sampled from $p_{\inn}$. We refer to one such complete execution of the circuit as a \emph{run}. Consequently, before the beginning of a run, all gates are expected to have their states from a previous run. This is formalized as follows. 

Let $p^t(\x)$ denote the distribution over the joint state of the circuit before execution step $t$, and let $p^0(\x)$ denote the distribution immediately before the beginning of a new run. Since $p^0$ is the distribution left at the end of the previous run, all gates are distributed according to the circuit computation:
\begin{equation}\label{eq_14}
p^0(\x)
=
\R(\x_{\nin}\mid\x_{\inn})p_{\inn}(\x_{\inn}).
\end{equation}
In particular, the states of the non-input gates are generally correlated with the input values from the previous run.

At execution step $t=0$, the input nodes are re-initialized by drawing a new input independently from $p_{\inn}$, while all non-input gates retain their previous states. Let
\begin{equation}
p_{\nin}^0(\x_{\nin})
=
\sum_{\x_{\inn}}p^0(\x_{\inn},\x_{\nin})
\end{equation}
denote the marginal distribution of the non-input gates before this re-initialization. The distribution immediately after the new input has been set is therefore
\begin{equation}\label{eq_15}
p^1(\x)
=
p_{\inn}(\x_{\inn})p_{\nin}^0(\x_{\nin}).
\end{equation}
Thus, the fresh input values are uncorrelated with the output states of non-input gates inherited from the previous run, while the marginal state of the non-input gates remains unchanged.

The remaining execution steps update the non-input gates one at a time. At step $t=1,\ldots,|V_{\nin}|$, gate $g_{t}$ is executed according to $\pi_{g_{t}} (\x_{g_{t}}\mid\x_{\pa(g_{t})})$ while the states of all other gates remain fixed. Consequently, after execution step $t$, the gates $g_1,\ldots,g_{t}$ have been updated during the current run, whereas $g_{t+1},\ldots,g_{|V_{\nin}|}$ still retain their states from the previous run. After the final execution step,
\begin{equation}
p^{|V_{\nin}|+1}(\x)
=
\R(\x_{\nin}\mid\x_{\inn})p_{\inn}(\x_{\inn})
=
p^0(\x),
\end{equation}
where the final equality denotes equality of distributions from one run to the next. The circuit is then ready for another run, beginning again with input re-initialization. 

Having specified the circuit dynamics, we now account for the mismatch cost incurred during a run. We begin with the re-initialization of the input nodes and then turn to the subsequent execution of the non-input gates.

\subsection{Mismatch cost of re-initializing the inputs}\label{sec3b}

During the input-re-initialization step, the input nodes are overwritten with newly sampled values, while the states of all non-input nodes remain unchanged. In this step the subsystem  $\x_{\inn}$ evolves independently of $\x_{\nin}$, with $\x_{\nin}$ held fixed. For such a subsystem process, the prior has product form~\cite{wolpert2019stochastic,yadav2025minimal},
\begin{equation}
    q^0(\x_{\inn},\x_{\nin})=q^0_{\inn}(\x_{\inn})q^0_{\nin}(\x_{\nin}).
\end{equation}
Since the re-initialization replaces the input state by a fresh sample from $p_{\inn}$, independently of the state of the non-input nodes, this prior evolves to
\begin{equation}
\qqt^0(\x_{\inn},\x_{\nin})=p_{\inn}(\x_{\inn})q^0_{\nin}(\x_{\nin}).
\end{equation}

Let $\MMC_{ow}(p_{\inn})$ denote the mismatch cost associated with re-initializing the inputs for input distribution $p_\inn$, 
\begin{equation}
    \MMC_{ow}(p_{\inn}) = D(p^0|q^0)-D(p^1|\qqt^0).
\end{equation}
Using Eqs.~\eqref{eq_14} and~\eqref{eq_15}, we obtain
\begin{equation}\label{eq:overwriting_mmc}
    \MMC_{ow}(p_{\inn}) = \I_0(\xx_{\inn};\xx_{\nin})+D(p_{\inn}|q^0_{\inn}), 
\end{equation}
where $\I_0(\xx_{\inn};\xx_{\nin})$ is the mutual information between input nodes and non-input gates before re-initializing the input nodes in the current run.  Because the states stored in the non-input nodes were generated during the previous execution of the circuit, the input and non-input nodes are in a correlated state $p^0(\x)$. After re-initialization of inputs by a fresh, independent sample from $p_{\inn}$, that mutual information is zero; $\I_1(\xx_{\inn};\xx_{\nin})=0.$ This loss of correlation gives rise to the mutual-information contribution in Eq.~\eqref{eq:overwriting_mmc}. The second term, $D(p_{\inn}|q^0_{\inn})$, is the mismatch between the actual distribution from which the new inputs are sampled and the input marginal of the EP-minimizing prior for the re-initialization process.

% {\color{red} Throughout the complexity-theoretic analysis below, the input-loading implementation and its least-dissipative input prior $q^0_{\inn}$ are held fixed as part of the thermodynamic circuit model. We vary $p_{\inn}$ as the operating distribution of fresh inputs presented to that fixed implementation; in particular, the supremum over $p_{\inn}$ introduced below does not re-optimize $q^0_{\inn}$ separately for each operating distribution.}

\subsection{Mismatch cost of gate-by-gate or layer-by-layer implementation of the
circuit}\label{sec:3c}

% Additionally, $q^0(\x)$ denotes the prior associated with the process of
% overwriting the inputs. When $\x_{g_t}$ is updated based on the values of its parents
% $\x_{\pa(g_t)}$, the rest of the nodes are unchanged. This makes the update a subsystem process
% in which $\x_{g_t}$ and $\x_{\pa(g_t)}$ form the local subsystem being acted upon. The prior
% distribution therefore has the product form
% \begin{equation}\label{eq:before_prior}
%     q^t(\x) = q_{g_t,\pa(g_t)}(\x_{g_t}, \x_{\pa(g_t)})\, q_{\widetilde{g_t}}(\x_{\widetilde{g_t}}),
% \end{equation}
% where $\x_{\widetilde{g_t}}$ denotes all the joint state of all gates other than $g_t$ and its parents.

Since the circuit is run repeatedly, with the same topological ordering applied on every run,
the marginal distribution over any gate's output state remains identical across runs. In particular, the marginal distribution over the output state of gate
$g_t$, denoted $p_{g_t}(\x_{g_t}) := \sum_{\x_{V\setminus g_t}} p^0(\x)$, is the same across any time step in a run, and similarly for the marginal distribution over its input states,
$p_{\pa(g_t)}(\x_{\pa(g_t)}) := \sum_{\x_{V\setminus \pa(g_t)}} p^0(\x)$.

Before gate $g_t$ updates its output in the current run, $\x_{g_t}$ retains its state from the
previous run, while $g_t$'s input states, $\x_{\pa(g_t)}$, have already been updated using
the current run's input sample (since the parents precede $g_t$ in the topological order). As
established in Sec.~\ref{sec3a}, a state retained from a previous run carries no correlation with gates that have
already updated in the current run; in particular, $\x_{g_t}$ and $\x_{\pa(g_t)}$ are
independent at this point. The joint distribution over the input and output states of the gate
is therefore
\begin{equation}\label{eq_actual_before}
    p^t_{g,\pa(g)}(\x_{g_t}, \x_{\pa(g_t)}) = p_{g_t}(\x_{g_t})\, p_{\pa(g_t)}(\x_{\pa(g_t)}).
\end{equation}
After gate $g_t$ executes, its new state is sampled from $\pi_{g_t}(\x_{g_t} \mid
\x_{\pa(g_t)})$, which depends only on the (unchanged) parent state $\x_{\pa(g_t)}$. The joint
distribution over the input and output states of the gate therefore becomes
\begin{equation}\label{eq_actual_after}
    p^{t+1}_{g,\pa(g)}(\x_{g_t}, \x_{\pa(g_t)}) = \pi_{g_t}(\x_{g_t} \mid \x_{\pa(g_t)})\,
    p_{\pa(g_t)}(\x_{\pa(g_t)}).
\end{equation}

We use $q_{g,\pa(g)}(\x_g, \x_{\pa(g)})$ to denote the prior distribution associated with the
process of updating the output state of gate $g$ based on its input states $\x_{\pa(g)}$; the
discussion of islands, uniqueness, and full support underlying this prior is given in
App.~\ref{App_prior_discussion}. When gate $g_t$ changes its state based on its parents'
states, the associated local prior distribution evolves to
\begin{equation}
    \qqt_{g_t,\pa(g_t)}(\x_{g_t}, \x_{\pa(g_t)}) = \pi_{g_t}(\x_{g_t} \mid \x_{\pa(g_t)})\,
    q_{\pa(g_t)}(\x_{\pa(g_t)}).
\end{equation}

In App.~\ref{app:MMC_circuits} we show that the mismatch cost associated with the execution of
gate $g_t$ in the middle of a run, at time step $t$, is given by
\begin{widetext}
\begin{equation}
    \MMC_t = S_{t+1}(\xx) - S_t(\xx) + \I_{t+1}(\xx_{g_t}; \xx_{\pa(g_t)}) +
    D(p_{g_t}p_{\pa(g_t)} \| q_{g_t,\pa(g_t)}) -
    D(p_{\pa(g_t)} \| q_{\pa(g_t)}), \label{eq_24}
\end{equation}
\end{widetext}
where $\I_{t+1}(\xx_{g_t}; \xx_{\pa(g_t)})$ denotes the mutual information between $g_t$ and
$\pa(g_t)$ after step $t$. The quantities $S_t(\xx)$ and $S_{t+1}(\xx)$ are the entropies of
the full circuit state before and after step $t$, respectively. The derivation of
Eq.~\eqref{eq_24} is provided in App.~\ref{app:MMC_circuits} (Eq.~\eqref{eq_24app}).

% where $q_{\pa(g_t)}$ is the same island-weighting distribution appearing in
% Eq.~\eqref{eq:local_prior_factored} (parent states are unchanged by this step, so the
% weighting across islands is unchanged as well). Since the rest of the gates do not change when
% gate $g_t$ is updated, $q^t(\x)$ evolves to
% \begin{equation}\label{eq:after_prior}
%     \qqt^t(\x) = \qqt_{g_t,\pa(g_t)}(\x_{g_t}, \x_{\pa(g_t)})\, q_{\widetilde{g_t}}(\x_{\widetilde{g_t}}).
% \end{equation}

\begin{figure*}
    \centering
    \includegraphics[trim={0 16cm 0 0}, width=0.99\textwidth]{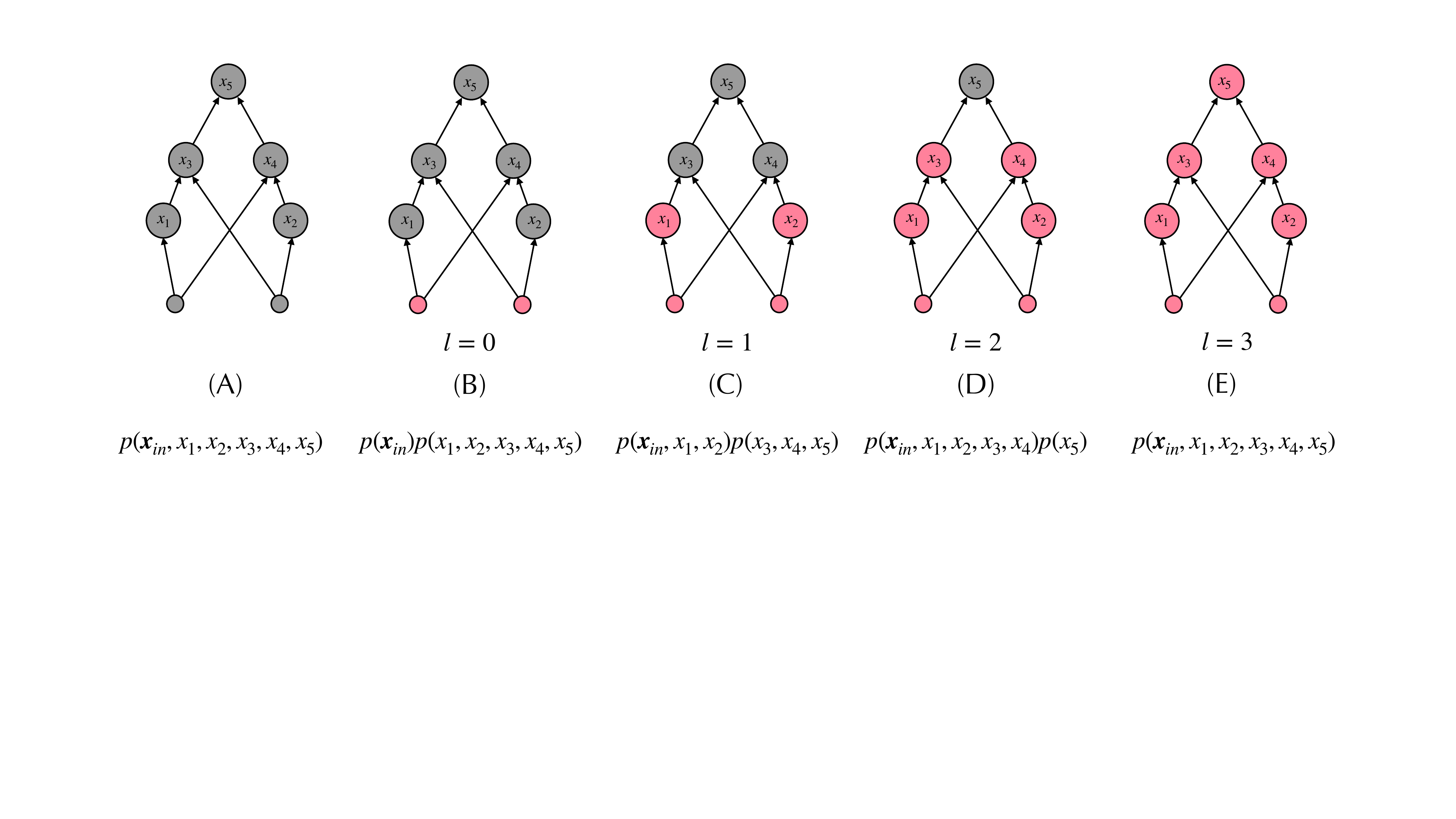}
    \caption{Layer-by-layer implementation of a circuit. (A) State of the circuit after a
    complete run. (B) Updating of input nodes for the next run, with values re-sampled from
    $p(\x_{\inn})$. (C) The first layer consists of gates $x_1$ and $x_2$. As they update
    together based on the newly sampled input values, they become correlated with the input
    nodes while losing correlation with the rest of the gates. Consequently, the joint
    distribution evolves to $p(\x_{\inn}, x_1, x_2)p(x_3, x_4, x_5)$. (D) Gates $x_3$ and $x_4$
    make up layer 2. As they update based on the new values of gates $x_1$ and $x_2$, they
    become correlated with them and the input nodes, while becoming independent of the state
    of $x_5$. The distribution then evolves to $p(\x_{\inn}, x_1, ..., x_4)p(x_5)$. Finally, in
    (E), gate $x_5$, constituting layer 3, is updated, and the distribution returns to its
    initial form as shown in (A).}
    \label{fig:illustration_circuits_layer}
\end{figure*}

% The mismatch cost of step $t$, i.e., of updating gate $g_t$, is
% \begin{equation}\label{eq:18}
%     \MMC_t = D(p^t\|q^t) - D(p^{t+1}\|\qqt^t).
% \end{equation}
% As a consequence of the island-respecting construction of $q_{g_t,\pa(g_t)}$ above,
% Eq.~\eqref{eq:18} is exactly the general, multi-island mismatch cost of
% Eq.~\eqref{eq:general_mismatch} applied to the local update of gate $g_t$, with islands
% indexed by the parent configuration $\x_{\pa(g_t)}$.

The total mismatch cost incurred over a full run -- including both the resetting of the inputs and the execution of all $|V_{\nin}|$ non-input gates -- is obtained by summing $\MMC_{ow}(p_{\inn})$, the mismatch cost of overwriting the inputs, together with Eq.~\eqref{eq_24} over the $|V_{\nin}|$ execution steps. This gives

\begin{widetext}
\begin{align}
    \MMC(C_n, p_{\inn}) &= \MMC_{ow}(p_{\inn}) + \sum_{t=1}^{T} \MMC_t \\
    &= D(p_{\inn}\|q_{\inn}) + \sum_{t=1}^{T} \I_{t+1}(\xx_{g_t}; \xx_{\pa(g_t)}) +
    \sum_{t=1}^{T} \left[D(p_{g_t}p_{\pa(g_t)}\|q_{g_t,\pa(g_t)}) -
    D(p_{\pa(g_t)}\|q_{\pa(g_t)})\right]. \label{eq_26}
\end{align}
\end{widetext}
A few remarks about Eq.~\eqref{eq_26} are in order. In particular, the expression for the MMC of a single step in Eq.~\eqref{eq_24} involves the entropy of the full circuit state, which is a global quantity. However, when these contributions are summed over all non-input gate updates, the intermediate entropy terms telescope and cancel. What remains are only boundary terms,
\begin{equation}
    S_1(\xx) - S_{T+1}(\xx),
\end{equation}
corresponding to the entropy of the joint circuit state immediately after the inputs are re-initialized and before any non-input gates are executed, and the entropy after all non-input gates have completed their updates. This boundary contribution cancels the mutual-information term appearing in the overwriting MMC in Eq.~\eqref{eq:overwriting_mmc}. Consequently, after adding the overwriting MMC, the total mismatch cost simplifies to Eq.~\eqref{eq_26}.

For deterministic circuits, the mutual-information change underlying this telescoping contribution also admits a direct interpretation in terms of correlations between a gate's value inherited from the previous run and the gates that have not yet been updated; see App.~\ref{app:future_mi}.

Equation~\eqref{eq_26} also yields the strictly positive lower bound stated in the title for deterministic circuits computing a nonconstant Boolean function through a non-input output gate. By Eq.~\eqref{eq_island2}, each gate-dependent KL difference in Eq.~\eqref{eq_26} can be written as
\begin{align}
&D(p_g p_{\pa(g)}\|q_{g,\pa(g)})
-D(p_{\pa(g)}\|q_{\pa(g)}) \nonumber\\
&\qquad =
\sum_{x_{\pa(g)}}p_{\pa(g)}(x_{\pa(g)})
D\!\left(
p_g\|q_{g\mid x_{\pa(g)}}
\right)
\ge 0.
\label{eq:positive_gate_term}
\end{align}
Thus every term in Eq.~\eqref{eq_26} is nonnegative. If $p_{\inn}\neq q_{\inn}$, then $D(p_{\inn}\|q_{\inn})>0$. If instead $p_{\inn}=q_{\inn}$, full support of $q_{\inn}$ implies full support of the operating input distribution. For a nonconstant deterministic output gate $g_{\rm out}$ this gives $S(X_{g_{\rm out}})>0$, while determinism gives $S(X_{g_{\rm out}}\mid X_{\pa(g_{\rm out})})=0$. Hence
\begin{equation}
I(X_{g_{\rm out}};X_{\pa(g_{\rm out})})=S(X_{g_{\rm out}})>0,
\end{equation}
and therefore
\begin{equation}
\MMC(C_n,p_{\inn})>0
\end{equation}
for every $p_{\inn}$ under these conditions. Since total entropy production is at least the mismatch cost, MMC is then a strictly positive lower bound on the thermodynamic cost of running the circuit.

It is straightforward to extend the analysis for a layer-by-layer implementation (see Fig.~\ref{fig:illustration_circuits_layer}). Based on a topological ordering of a DAG, it is possible to stratify the set of gates in a natural way into layers. Starting from the set of input nodes
\begin{equation}
    V_0 := \{ g \in V : \pa(g) = \emptyset \},
\end{equation}
we define the $l$-th layer recursively by
\begin{equation}
    V_l := \left\{ g \in V \setminus \bigcup_{k=0}^{l-1} V_k : \pa(g) \subseteq \bigcup_{k=0}^{l-1} V_k \right\}.
\end{equation}
This construction assigns each gate to the earliest layer in which all of its parents have already been evaluated, thereby inducing a valid feedforward layering of the circuit.
Let $\xx_l$ be the random variable denoting the joint state of gates in layer $l$ and let $\x_l$ denote a value of $\xx_l$. In a layer-by-layer implementation, all the gates in a layer are updated simultaneously. Let $d$ be the number of layers in the circuit. Then, analogous to Eq.~\eqref{eq_26}, the total mismatch cost of running all non-input layers is given by
\begin{align}\label{eq:MMC_layer2}
    \MMC_{ow} &+ \sum_{l = 1}^{d}\MMC_l = D(p_\inn||q_\inn) + \sum_{l = 1}^d \I_{l+1}(\xx_l; \xx_{\pa(l)}) \\ &+ \sum_{l = 1}^d \left[D\left(p_l p_{\pa(l)}||q_{l, \pa(l)}\right)\nonumber - D\left(p_{\pa(l)}||q_{\pa(l)}\right)\right], 
\end{align}
where $p_l$, $p_{\pa(l)}$, $q_{l, \pa(l)}$, and $q_{\pa(l)}$ are defined for layer $l$ and have their usual meaning as in Eq.~\eqref{eq_26}. 
Updating multiple gates simultaneously in a layer-by-layer approach, instead of updating gates sequentially one by one, represents a form of time-coarse graining. This is because grouping several updates into a single time step effectively reduces the temporal resolution of the system's dynamics. 

\begin{figure}
    \centering
    \includegraphics[width=0.99\linewidth]{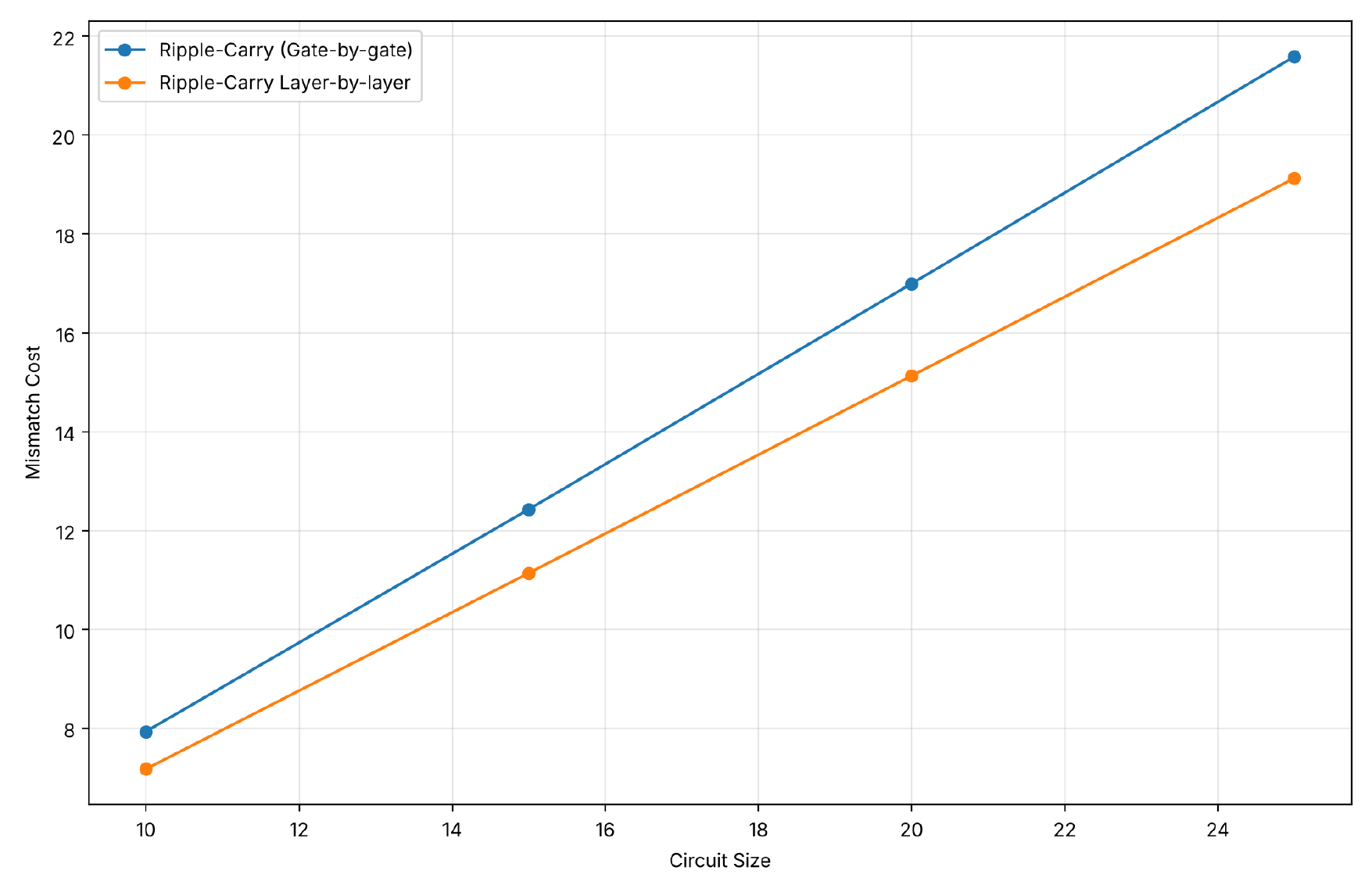}
    \caption{Comparison of layer-by-layer and gate-by-gate mismatch costs for the ripple-carry adder circuit family. Notably, the layer-by-layer mismatch cost remains consistently lower than the gate-by-gate mismatch cost. In this comparison, the input distribution is uniformly random for both implementations, and the priors of all gates or layers in the circuits are also uniform.}
    \label{fig:comparisoin_layer_gates}
\end{figure}

As demonstrated in~\cite{yadav2026entropy}, mismatch cost decreases under temporal coarse-graining of a fixed underlying process. Applied to circuits, this implies that when the layer-by-layer description is obtained by coarse-graining the same gate-by-gate physical process---so that the layer priors are those induced by that coarse-graining rather than independently chosen priors---the layer-by-layer mismatch cost is no greater than the mismatch cost of the corresponding gate-by-gate description. In Fig.~\ref{fig:comparisoin_layer_gates}, we compare the mismatch costs of layer-by-layer and gate-by-gate implementations of a ripple-carry adder. Gate-by-gate mismatch cost is computed using Eq.~\eqref{eq_26}, while the layer-by-layer mismatch cost is computed using Eq.~\eqref{eq:MMC_layer2}. As a consequence of the lowering in mismatch cost under time-coarse graining, the layer-by-layer mismatch cost is lower than the gate-by-gate mismatch cost across ripple-carry adder circuits with varying input lengths.

Additionally, note that as the computation in a circuit progresses---whether gate-by-gate or layer-by-layer---the distribution over the states cycles through a sequence of distributions (see Fig.~\ref{fig:illustration_circuits_layer}). Moreover, the phase at which the sequence of distributions is started does not impact the final mismatch cost, as long as the full cycle of execution is completed.

The sequence of gate implementations in a circuit can be modeled as a periodic process. In practical computational devices based on circuits, the execution of gates or layers is typically driven by a clock~\cite{weste2015cmos}. From this perspective, the process is periodic, with gates executed at each clock cycle, sequentially leading to the completion of the full circuit. In Appendix~\ref{app:circuit_periodic}, we present the MMC calculation for this clock-driven model of circuit execution. This approach yields the same expression for the MMC as given in Eq.~\eqref{eq_26}.

In the next section, we use Eq.~\eqref{eq_26} to calculate the mismatch cost of various circuit families and broadly investigate how circuit structure affects their mismatch cost. In particular, we define mismatch cost complexity, analogous to the size and depth complexity of a circuit family. We will then establish results that relate mismatch cost complexity to size complexity.

\begin{figure}
    \centering
    \includegraphics[width=0.99\linewidth]{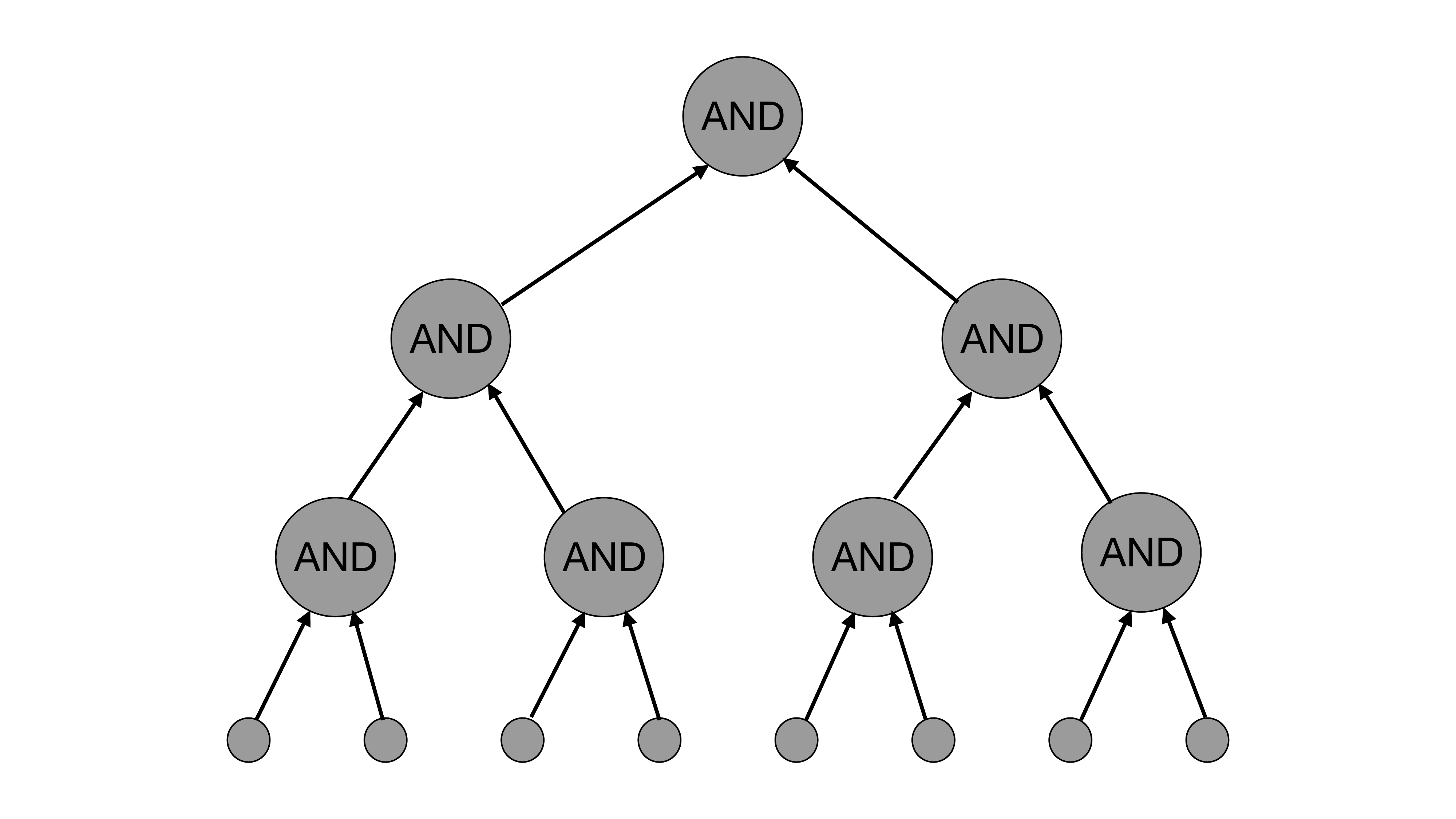}
    \caption{Binary tree of AND gates. 
    The size of the tree grows linearly with $n$, whereas its depth grows as $\log n$.}
    \label{fig:AND_tree_illustration}
\end{figure}

\section{Mismatch cost complexity of circuit families}\label{sec:4}

Having derived the mismatch cost of a circuit in Sec.~\ref{sec:3}, we now ask how this cost scales across a circuit family as the input size grows, and how that scaling relates to conventional circuit resources such as size and depth. We first define mismatch cost complexity relative to a fixed gate basis with priors, and then derive a general upper bound relating it to the number and types of gates in the circuit. Homogeneous and heterogeneous prior bases arise as useful special cases of this general bound.

\subsection{Definition and general relationship to circuit size}\label{sec:4a}

For a gate basis $\basis$, we augment each gate type $b\in\basis$ with its associated conditional island priors $q_{b\mid\pa(b)}$, and denote the resulting \emph{basis with priors} by
\begin{equation}
    \widetilde{\basis}
    :=
    \{(b,q_{b\mid\pa(b)}):b\in\basis\}.
\end{equation}
Equivalently, these conditional priors may be represented using the joint bookkeeping distributions $q_{b,\pa(b)}$ introduced in Sec.~\ref{sec:3c}. The conditional priors are the physically relevant quantities: as discussed in App.~\ref{App_prior_discussion}, they are unique and have full support within each parent-state island.

For a circuit $C_n$ with input distribution $p_{\inn}$, let $\MMC(C_n,p_{\inn})$ denote the total mismatch cost incurred during one complete run, as given by Eq.~\eqref{eq_26}. We define the worst-case mismatch cost of $C_n$ by
\begin{equation}
    \MMC(C_n)
    :=
    \sup_{p_{\inn}} \MMC(C_n,p_{\inn}).
\end{equation}

\begin{definition}[\textbf{Mismatch cost complexity}]
    Given a basis with priors $\widetilde{\basis}$, a circuit family $\{C_n\}_{n\in\N}$ is said to have mismatch cost complexity $m(n)$ if
    \begin{equation}
        \MMC(C_n)=O(m(n)).
    \end{equation}
\end{definition}

Analogously, we define the associated mismatch cost complexity class.

\begin{definition}[\textbf{Mismatch cost complexity class}]
    Let $m:\N\to\mathbb{R}^+$. For a fixed basis with priors $\widetilde{\basis}$, let $\MMClass{m}$ denote the class of functions $f$ for which there exists a circuit family $\C=\{C_n\}_{n\in\N}$ computing $f$ and having mismatch cost complexity $m(n)$.
\end{definition}

The class $\MMClass{m}$ is therefore always understood relative to the specified basis with priors $\widetilde{\basis}$. A given logical gate can be realized by distinct physical processes, with different least-dissipative priors and different residual EP. Accordingly, $\MMClass{m}$ does not classify the total EP of a Boolean function independently of its physical realization. Once $\widetilde{\basis}$ is fixed, however, together with the input distribution, the topology determines the distributions encountered by individual gates and therefore their contributions to Eq.~\eqref{eq_26}.

As discussed in App.~\ref{App_prior_discussion}, for each fixed parent state $x_{\pa(g)}$, the conditional prior $q_{g\mid\pa(g)}(x_g\mid x_{\pa(g)})$ is unique and has full support over the gate-output state space. Therefore,
\begin{equation}
    q^{\min}_{g\mid\pa(g)}
    :=
    \min_{x_g,x_{\pa(g)}}
    q_{g\mid\pa(g)}(x_g\mid x_{\pa(g)})
\end{equation}
is strictly positive. The same holds for the prior over the circuit inputs,
\begin{equation}
    q^{\min}_{\inn}
    :=
    \min_{x_{\inn}}
    q_{\inn}(x_{\inn}).
\end{equation}

We now show that full support of these priors yields a finite gate-wise upper bound on the mismatch cost, and that a uniform version of this bound relates mismatch cost complexity directly to circuit size.

\begin{theorem}\label{th:1}
Let $C_n$ be a Boolean circuit evaluated in the gate-by-gate model, with input prior $q_{\inn}$ and local gate priors $q_{g,\pa(g)}$ for every non-input gate $g$. Then, for every input distribution $p_{\inn}$,
\begin{equation}\label{eq:general_mmc_bound}
    \MMC(C_n,p_{\inn})
    \le
    \log\frac{1}{q^{\min}_{\inn}}
    +
    \sum_{t=1}^{T}
    \log\frac{1}{q^{\min}_{g_t\mid\pa(g_t)}}.
\end{equation}
In particular, if the circuit family satisfies
\begin{align}
    \log\frac{1}{q^{\min}_{\inn}}
    &\le nK_{\inn},\\
    \log\frac{1}{q^{\min}_{g_t\mid\pa(g_t)}}
    &\le K
    \qquad
    \text{for all }g_t\in V_{\nin},
\end{align}
for constants $K_{\inn}$ and $K$ independent of $n$, then
\begin{equation}\label{eq:general_mmc_bound_linear}
    \MMC(C_n)
    \le
    |C_n|K+nK_{\inn}.
\end{equation}
\end{theorem}

The proof is given in App.~\ref{proof2}. Theorem~\ref{th:1} immediately relates MMC complexity to circuit size. If $s(n)$ denotes the size complexity of the family, then Eq.~\eqref{eq:general_mmc_bound_linear} implies
\begin{equation}\label{eq:40}
    m(n)=O\!\bigl(s(n)+n\bigr),
\end{equation}
with constants determined by $K$ and $K_{\inn}$. By Eq.~\eqref{eq:40}, if the size complexity $s(n)$ of a circuit family grows as a polynomial function of the input size, then the upper bound on the mismatch cost complexity for any homogeneous prior basis also scales polynomially with the input size. In other words, any Boolean function that can be implemented by a polynomial-size circuit can also be realized by a circuit family whose mismatch cost complexity does not exceed polynomial growth. This observation leads to the following corollary:

\begin{corollary}\label{corollary1}
For any fixed finite universal Boolean basis $\basis$ and any associated homogeneous prior basis $\basisg$,
    \begin{equation}
        \Poly \subseteq \MMClass{\poly} 
    \end{equation}
\end{corollary}

It remains yet to be determined whether the converse holds---that is, whether any function implementable by a circuit family with a mismatch cost that grows no faster than polynomial can also be realized by a circuit family whose size complexity is polynomially bounded, and thus whether $\Poly = \MMClass{\poly}$. 

\subsection{Homogeneous product priors}\label{sec:4b}

The general bound in Theorem~\ref{th:1} simplifies considerably under additional assumptions on the priors. Suppose that:

\begin{enumerate}
    \item for every non-input gate $g\in V_{\nin}$, the local prior factorizes as
    \begin{equation}
        q_{g,\pa(g)}(x_g,x_{\pa(g)})
        =
        q_g(x_g)q_{\pa(g)}(x_{\pa(g)});
    \end{equation}

    \item all non-input gates share a common marginal prior $q_\star$, i.e.,
    \begin{equation}
        q_g=q_\star
        \qquad
        \text{for all }g\in V_{\nin};
    \end{equation}

    \item the input prior factorizes identically over the $n$ input bits,
    \begin{equation}
        q_{\inn}(x_{\inn,1},\ldots,x_{\inn,n})
        =
        \prod_{k=1}^n q_i(x_{\inn,k}).
    \end{equation}
\end{enumerate}

The first assumption implies
\begin{equation}
    q_{g\mid\pa(g)}(x_g\mid x_{\pa(g)})=q_g(x_g),
\end{equation}
so the gate-wise constants depend only on the marginal gate prior. The second assumption makes this constant common to all non-input gates, while the third ensures that the input-prior contribution grows linearly with the number of input bits. Consequently, as derived in App.~\ref{proof2},
\begin{equation}\label{eq:rest38}
    \MMC(C_n)
    \le
    |C_n|\log\frac{1}{q_\star^{\min}}
    +
    n\log\frac{1}{q_i^{\min}}.
\end{equation}

\begin{figure}
    \centering
    \includegraphics[width=0.99\linewidth]{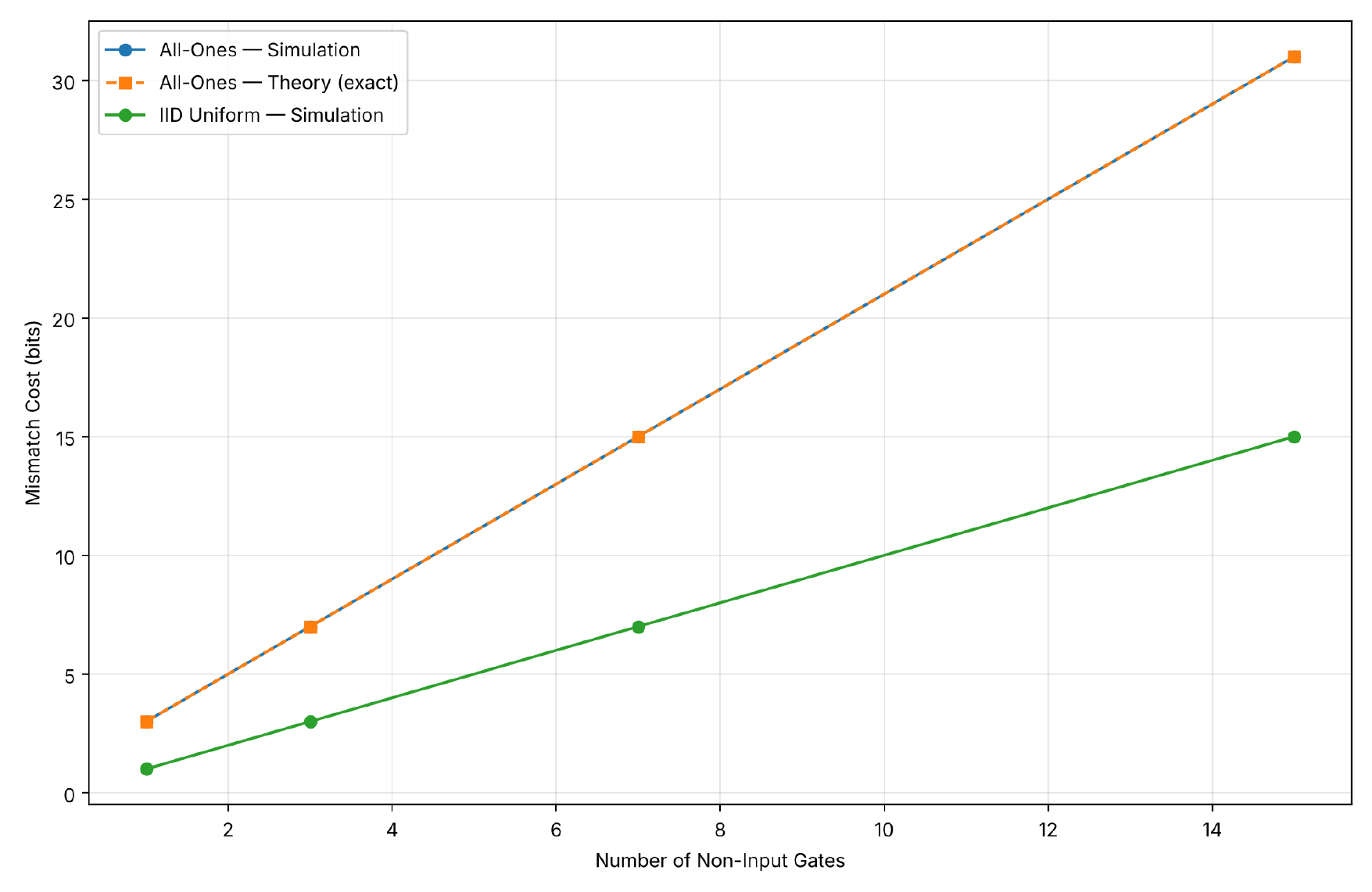}
    \caption{Saturating the Upper Bound on MMC for the AND Binary Tree Circuit Family. The solid blue line represents the total mismatch cost obtained by using Eq.~\eqref{eq_26} and a prior that is uniform for an input distribution consisting entirely of all ones, matching exactly with the orange dashed line, which corresponds to the theoretical upper bound given by Eq.~\eqref{eq:MMC_C_n}. In contrast, when the input distribution is a uniform distribution, the actual total mismatch cost of the AND binary tree remains consistently lower than the theoretical upper bound: the solid green line depicts the total mismatch cost for a uniform input distribution.}
    \label{fig:saturating_MMC_AND}
\end{figure}

This linear upper bound can be approached by suitable circuit families under suitable worst-case input distributions. As an example, consider the family $C_n$ of binary AND trees. Since a binary AND tree with $n$ input bits has $|C_n|=n-1$ non-input gates, Eq.~\eqref{eq:rest38} becomes
\begin{equation}\label{eq:AND_binar_upper_bound}
    \MMC(C_n)
    \le
    (n-1)\log\frac{1}{q_\star^{\min}}
    +
    n\log\frac{1}{q_i^{\min}}.
\end{equation}
As shown in Fig.~\ref{fig:saturating_MMC_AND}, when the input distribution is concentrated on the all-ones string, the actual mismatch cost approaches the theoretical upper bound. For other input distributions, such as the uniform distribution, the bound is not saturated and the actual mismatch cost remains lower.

\subsection{Heterogeneous gate priors}\label{sec:4c}

We next relax the homogeneous-prior assumption and allow different gate types to have different priors. Suppose that every occurrence of gate type $b\in\basis$ is assigned the same conditional prior $q_{b\mid\pa(b)}$, but that these priors may differ across gate types. We refer to such a setting as \emph{heterogeneous}. Define
\begin{equation}
    q^{\min}_{b\mid\pa(b)}
    :=
    \min_{x_b,x_{\pa(b)}}
    q_{b\mid\pa(b)}(x_b\mid x_{\pa(b)}),
\end{equation}
and
\begin{equation}
    K_b
    :=
    \log\frac{1}{q^{\min}_{b\mid\pa(b)}}.
\end{equation}
By the same full-support argument used above, $K_b$ is finite for every $b\in\basis$.

\begin{theorem}\label{th:3}
Let $\basis$ be a finite basis of logic-gate types, and suppose that every occurrence in $C_n$ of gate type $b\in\basis$ is assigned the same conditional prior $q_{b\mid\pa(b)}$. If $\#_b(n)$ denotes the number of gates of type $b$ in $C_n$, then
\begin{equation}\label{eq:th:3}
    \MMC(C_n)
    \le
    \sum_{b\in\basis}
    \#_b(n)K_b
    +
    \log\frac{1}{q^{\min}_{\inn}}.
\end{equation}
If in addition
\begin{equation}
    \log\frac{1}{q^{\min}_{\inn}}
    \le nK_{\inn},
\end{equation}
then
\begin{equation}\label{eq:th3_linear}
    \MMC(C_n)
    \le
    \sum_{b\in\basis}
    \#_b(n)K_b
    +
    nK_{\inn}.
\end{equation}
\end{theorem}

The theorem follows directly from Eq.~\eqref{eq:general_mmc_bound}: each occurrence of gate type $b$ contributes at most $K_b$, and summing over all gate types gives Eq.~\eqref{eq:th:3}. If the input prior factorizes across input bits,
\begin{equation}
    q_{\inn}(x_{\inn})
    =
    \prod_{i=1}^{n}
    q_{\inn,i}(x_{\inn,i}),
\end{equation}
then
\begin{equation}
    q^{\min}_{\inn}
    =
    \prod_{i=1}^{n}
    q^{\min}_{\inn,i}
    \ge
    \left(q^{\min}_{\inn,\ast}\right)^n,
\end{equation}
where
\begin{equation}
    q^{\min}_{\inn,\ast}
    :=
    \min_i q^{\min}_{\inn,i}.
\end{equation}
Setting $K_{\inn}:=\log(1/q^{\min}_{\inn,\ast})$ yields the linear input contribution in Eq.~\eqref{eq:th3_linear}.

Theorem~\ref{th:3} shows that for a heterogeneous prior basis, the natural analogue of circuit size is a \emph{weighted gate count}: each occurrence of gate type $b$ contributes a weight $K_b$. Thus, two circuits with the same total number of gates can nevertheless have different mismatch-cost bounds because their gate compositions differ. In the homogeneous case, where all $K_b$ are equal, this weighted count reduces to an ordinary size-dependent bound.

The product-prior assumption can also be imposed in the heterogeneous setting. If, for each gate type $b\in\basis$,
\begin{equation}
    q_{b,\pa(b)}(x_b,x_{\pa(b)})
    =
    q_b(x_b)q_{\pa(b)}(x_{\pa(b)}),
\end{equation}
then
\begin{equation}
    q_{b\mid\pa(b)}=q_b,
    \qquad
    K_b
    =
    \log\frac{1}{q_b^{\min}}.
\end{equation}
Thus, product factorization is not required for Theorem~\ref{th:3}; it is needed only if one wishes to express the type-dependent constants in terms of marginal gate priors rather than conditional island priors.

\begin{figure}
    \centering
    \includegraphics[width=0.99\linewidth]{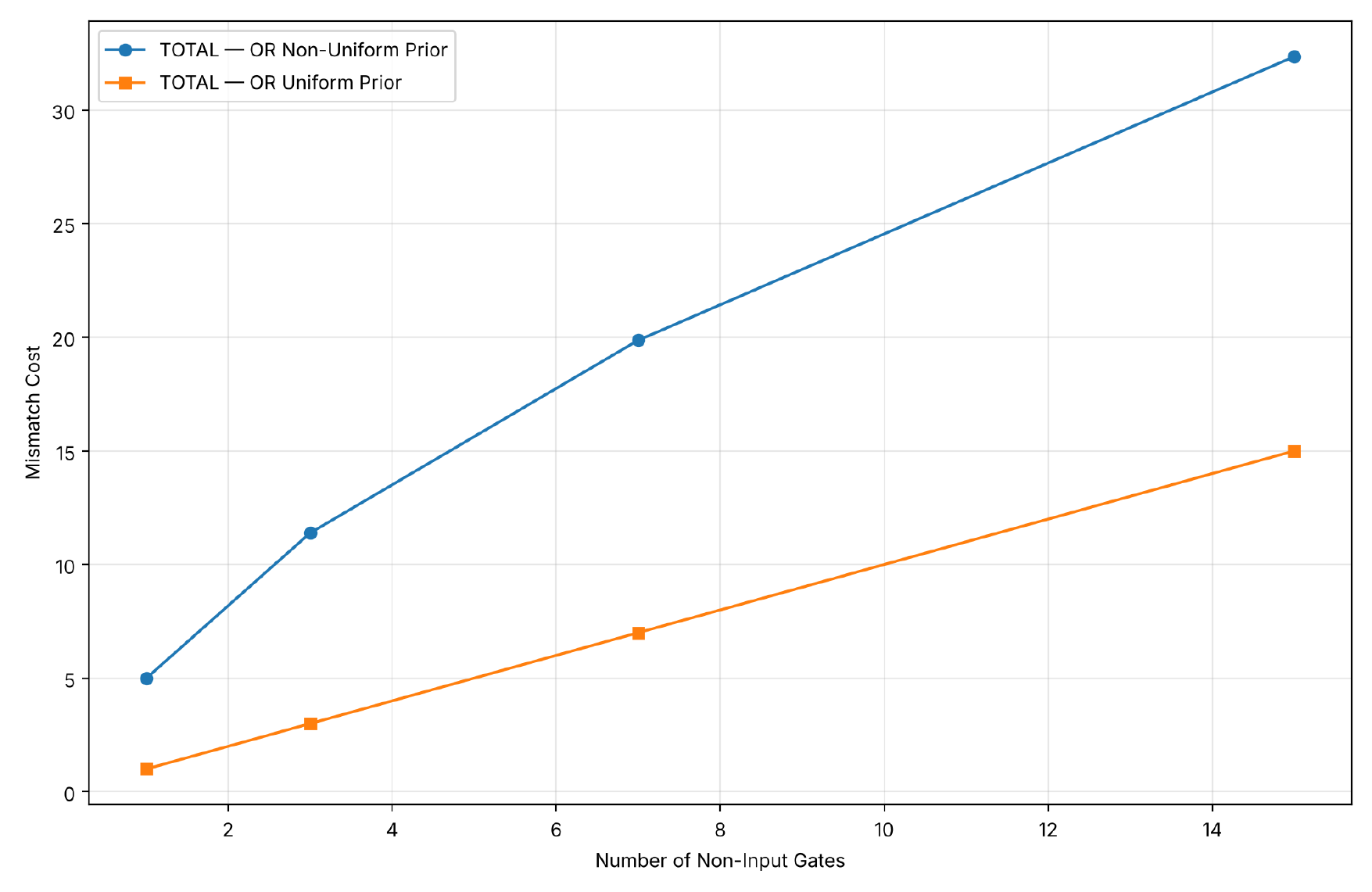}
    \caption{Mismatch cost scaling for a circuit family with homogeneous (orange) vs. heterogeneous priors (blue). The circuit family that we consider here consists of binary trees made up of AND and OR gates; each layer has one OR gate and rest of the gates are AND. The solid blue curve shows the mismatch cost when the prior of OR gates differs from the uniform prior of AND gates (heterogeneous priors). For the case where AND and OR gates have different prior distributions, the mismatch cost is not linear with circuit size. In contrast, the orange line represents the mismatch cost for the same circuit family with identical priors for OR and AND gates (homogeneous priors), and the mismatch cost scales linearly with the circuit size.}
    \label{fig:heterogeneous_prior}
\end{figure}

Figure~\ref{fig:heterogeneous_prior} illustrates one circuit family in which a changing composition of gate types, together with heterogeneous priors, produces a nonlinear relationship between total mismatch cost and total gate count. More generally, the content of Theorem~\ref{th:3} is that mismatch cost is controlled by the weighted composition $\{\#_b(n)\}_{b\in\basis}$ of the circuit family, rather than by its total size alone.

\begin{figure}
    \centering
    \includegraphics[width=0.99\linewidth]{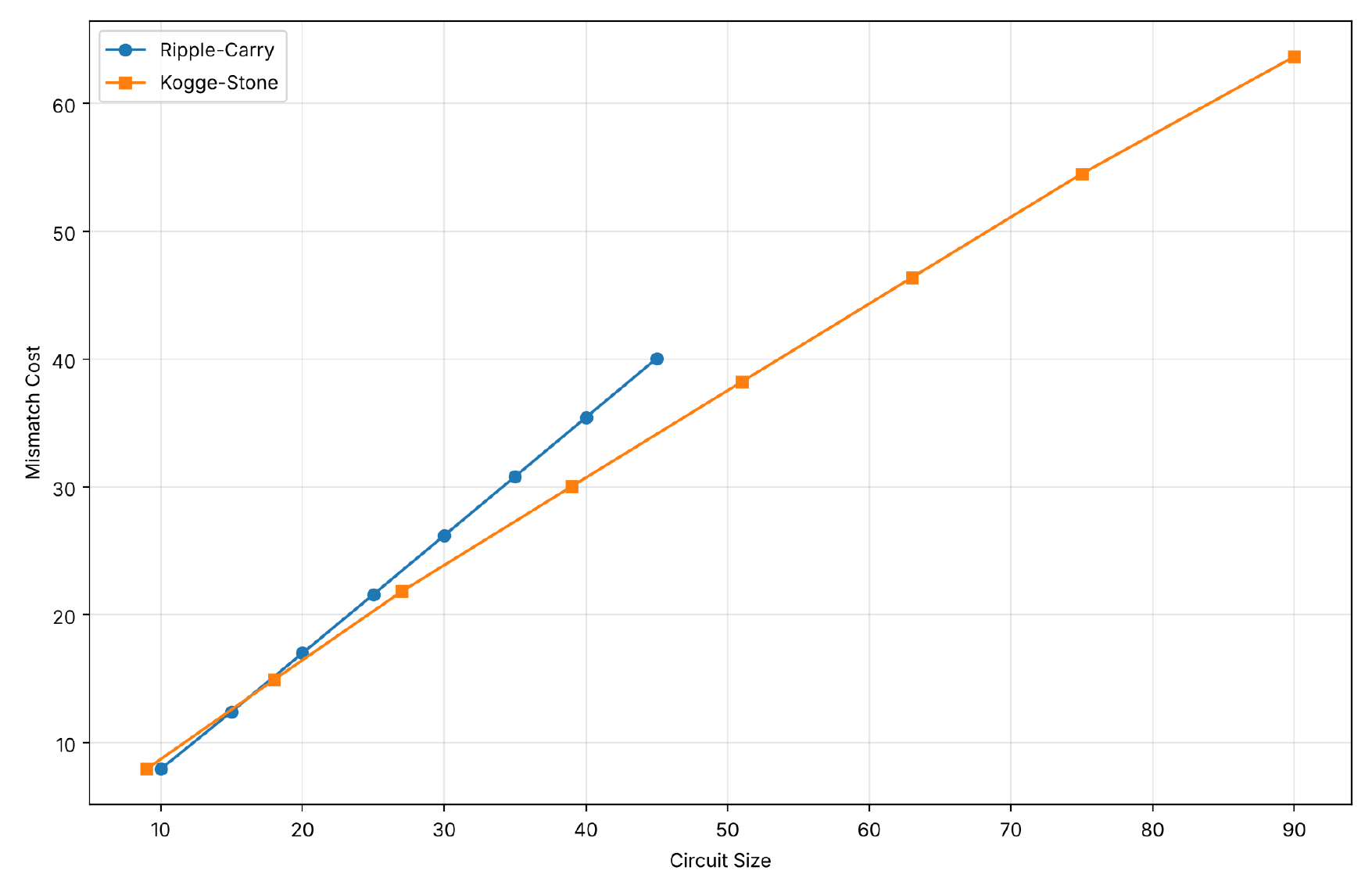}
    \caption{The MMC of two different circuit families implementing the addition function, which takes two integers in the standard $n$-bit representation and returns their sum in the standard $(n+1)$-bit representation.}
    \label{fig:ADD_sequential_parallel}
\end{figure}

\begin{figure}
    \centering
    \includegraphics[width=0.99\linewidth]{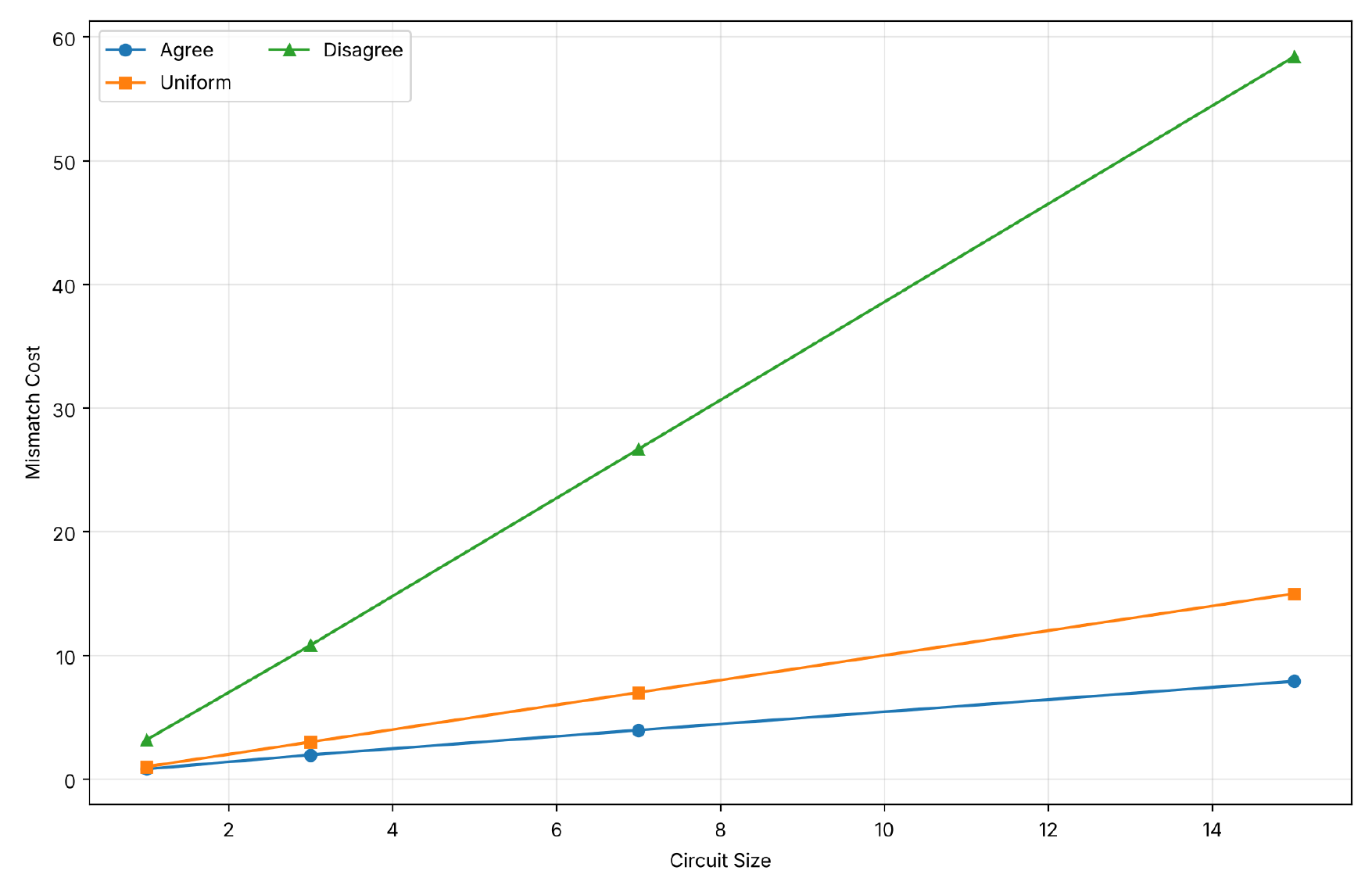}
    \caption{Mismatch cost for a family of binary-tree AND circuits under different gate priors.
    For a uniform input distribution, the evaluation of the circuit induces a joint distribution over output $\x_g$ and inputs $\x_{\pa(g)}$ of each gate $g$. Just before gate $g$ updates, its output still holds its value from the previous run, which is independent of its new inputs. The figure shows the total mismatch cost for three choices of the gate prior $q_g(\x_g, \x_{\pa(g)})$: (blue) a prior `agreeing' with this pre-update state, $q_{g|\pa(g)}(\x_g \mid \x_{\pa(g)}) = p_g(\x_g)$; (orange) a uniform prior over the joint input–output states; (green) a prior that assigns 97\% probability to the output opposite to that of an ideal AND gate (`disagreeing' with the ideal AND behavior). The comparison illustrates how agreement or disagreement between the prior and the state of the gate before it updates impacts the total mismatch cost.}
    \label{fig_9}
\end{figure}
\begin{figure}
    \centering
    \includegraphics[width=0.99\linewidth]{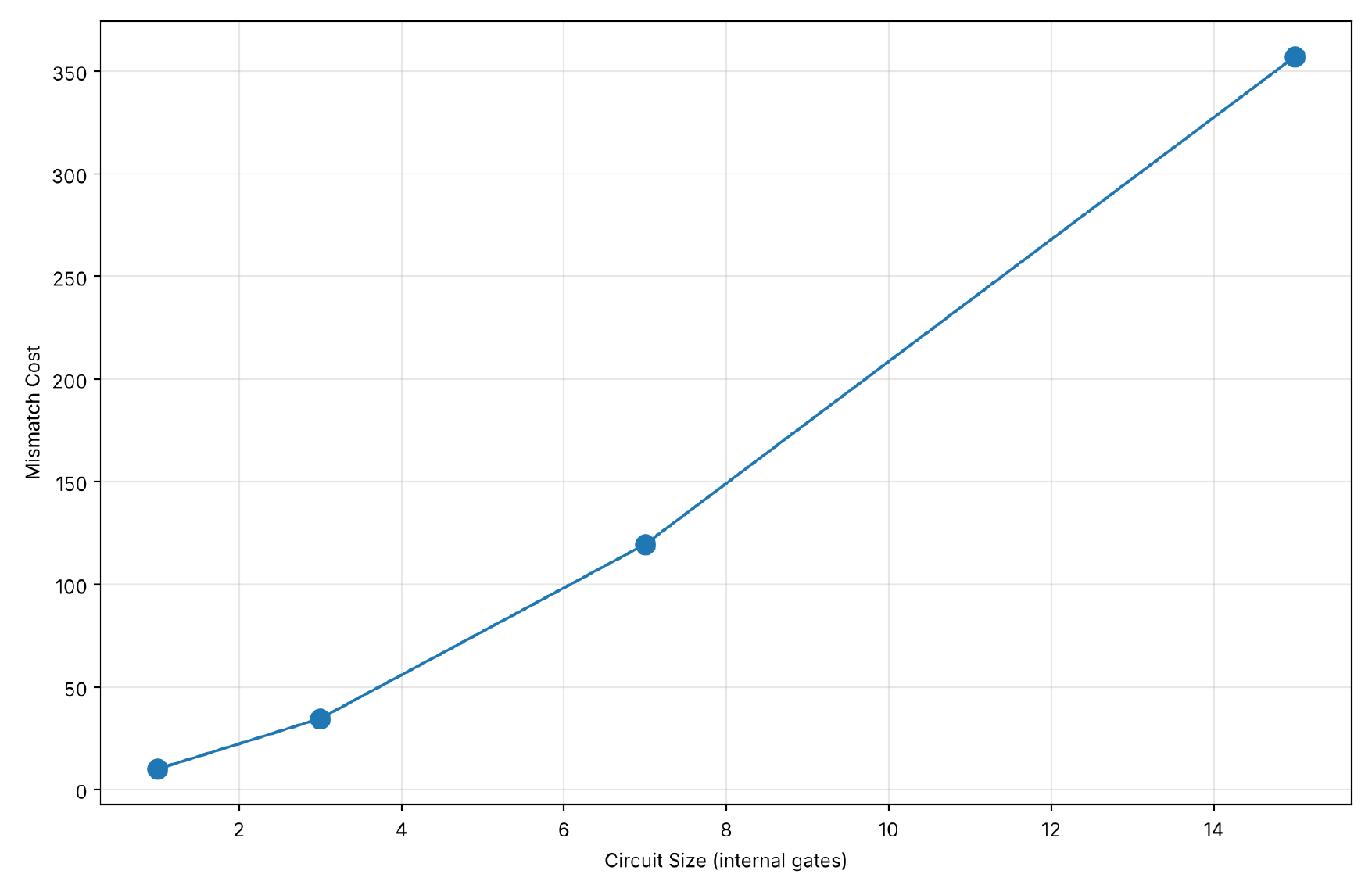}
    \caption{Mismatch cost of a mixed binary tree circuit family. This figure illustrates that the linear cost–size relation need not hold for circuit families in the finite input regime. The circuit family is constructed as follows: starting with a depth-1 tree which consists of a CONST gate that always outputs zero, the depth-2 tree is made of AND gates, depth-3 tree is made from NOR gates and lastly, depth 4 tree is a tree of NAND gates. All gates are evaluated under the same prior: inputs are assumed independent and uniformly distributed, so that $q_{\pa(g)}(\x_\pa(g)) = \tfrac{1}{4}$ for all $\x_\pa(g) \in \{0,1\}^2$, and $q_{g|\pa(g)}(\x_g \mid \x_\pa(g))~=~\begin{cases}
    1 - \varepsilon & \text{if } g = x \land y, \\
    \varepsilon & \text{otherwise},
    \end{cases}$, with $\varepsilon = 10^{-12}$. This means that the prior strongly favors AND behavior, and deviations from AND incur mismatch cost. Jointly, $q_{\pa(g)}$ and $q_{g|\pa(g)}$ determine the prior $q_{g,~\pa(g)}(\x_g,~\x_{\pa(g)})~=~q_{g|\pa(g)}(\x_g~\mid~\x_\pa(g))q_{\pa(g)}(\x_\pa(g))$.}
    \label{fig_10}
\end{figure}

\begin{figure}
    \centering
    \includegraphics[width=0.99\linewidth]{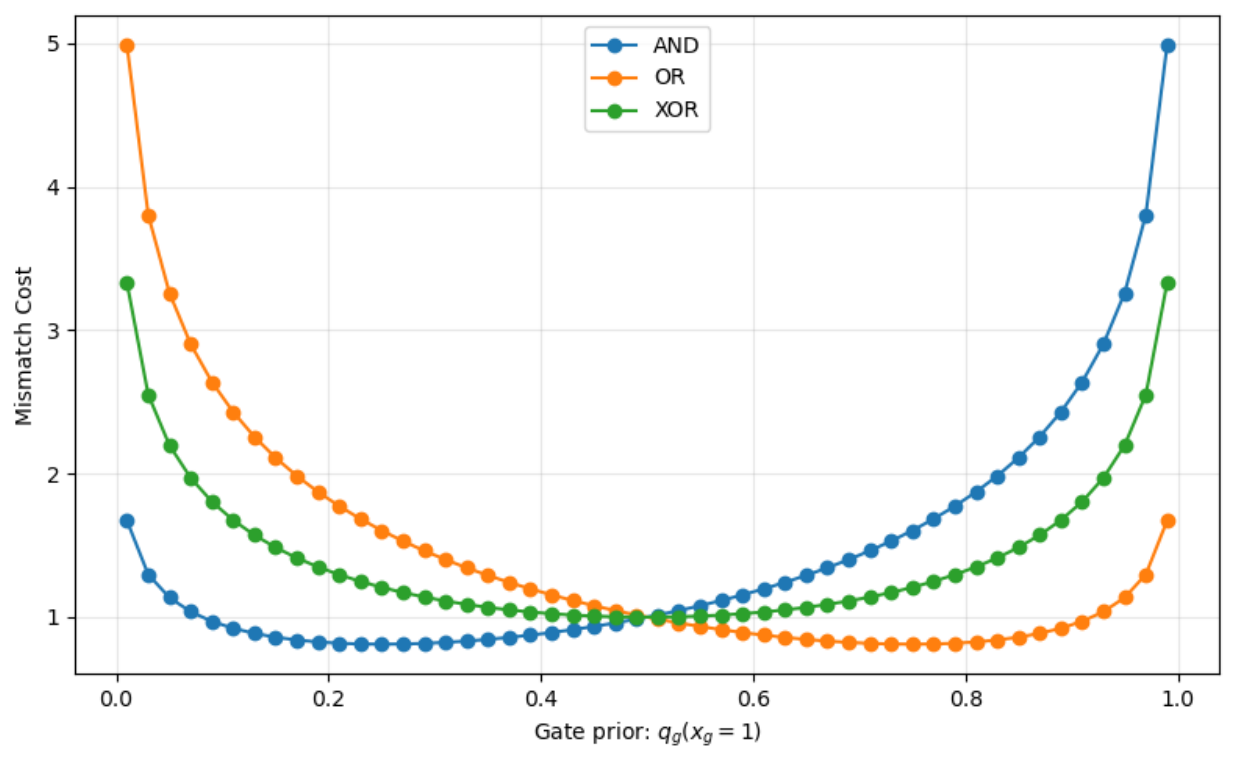}
    \caption{Mismatch cost of individual gates as a function of their prior distributions. The input distribution to the gate is uniform. The x-axis represents the support of prior distribution at $x_g = 1$.}
    \label{fig:11}
\end{figure}

\begin{figure}
    \centering
    \includegraphics[width=0.99\linewidth]{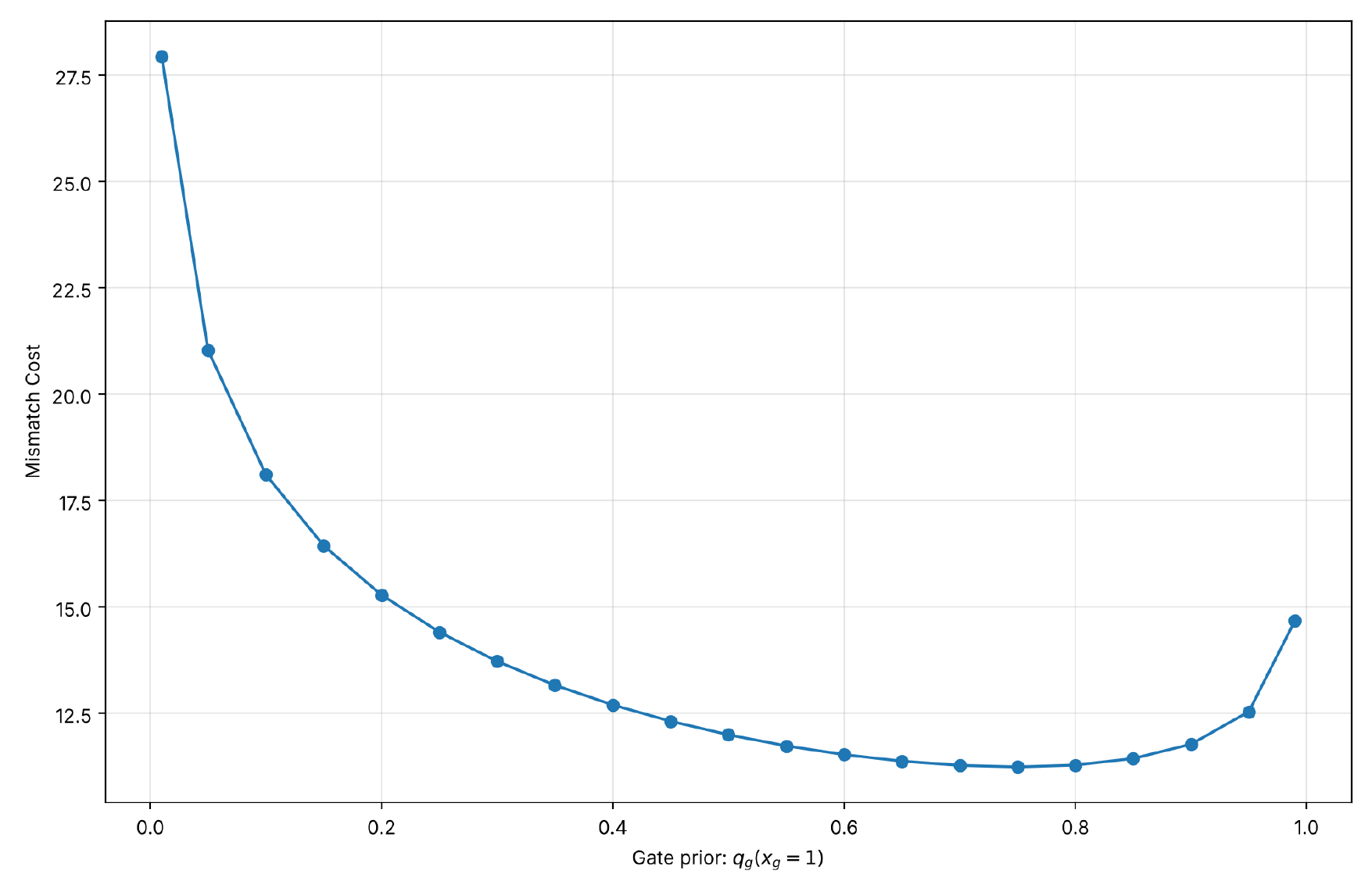}
    \caption{Total mismatch cost of a circuit with 4 layers, each containing two AND gates and one OR gate. The prior distribution for AND gates is fixed and uniform, while the prior of OR gates is varied. The plot shows how the total mismatch cost changes with the OR gate prior.}
    \label{fig:12}
\end{figure}

\section{Circuit design and resource trade-offs}\label{sec:5}

The results of Sec.~\ref{sec:4} establish mismatch cost as a circuit resource whose scaling can be related to, but is not determined solely by, conventional measures such as circuit size. This raises a natural design question: among different circuit families that implement the same Boolean function, how does minimizing mismatch cost compare with minimizing standard resources such as size or depth? In this section, we illustrate that these optimization criteria can distinguish circuit architectures in different ways.

\subsection{Different circuit families implementing the same function}\label{sec:5a}

A Boolean function can generally be implemented by many different circuit families, whose size and depth complexities may differ substantially. Standard circuit complexity compares such implementations according to these resources. Once a basis with priors and an input distribution are specified, mismatch cost provides an additional criterion by which the same implementations can be compared.

As an example, consider the Boolean function $\ADD^n$, which computes the sum of two $n$-bit binary numbers, as discussed in Sec.~\ref{sec:2a}. Two standard circuit families implementing this function are the ripple-carry adder (RCA) and the carry look-ahead adder (CLA). The RCA has linear size and linear depth, whereas the CLA reduces the depth to $O(\log n)$ at the cost of increasing the size to $O(n\log n)$.

In Fig.~\ref{fig:ADD_sequential_parallel}, we compare the mismatch costs of these two circuit families under the same input distribution using for each gate a prior equal to the distribution of its output immediately before execution. For the choices used in the figure, the MMC of both families grows approximately linearly with circuit size, but the RCA incurs a smaller total mismatch cost than the CLA as the input size increases. Thus, although the CLA is preferable from the standpoint of depth, the RCA is preferable with respect to mismatch cost under these modeling choices. This example illustrates that different resource measures can rank circuit implementations of the same function differently.

\subsection{Minimizing conventional complexity need not minimize mismatch cost}\label{sec:5c}

More generally, minimizing circuit size need not minimize mismatch cost. The reason is that
MMC depends not only on the number of gates, but also on the distributions encountered by
those gates and on how well those distributions are matched to the corresponding gate priors.
Consequently, circuits of the same size can have different mismatch costs, and a larger circuit
can in some cases have lower MMC than a smaller implementation of the same function.

The dependence on the prior basis is illustrated in Fig.~\ref{fig_9}, where the same family
of binary-tree AND circuits is evaluated under several choices of gate prior. Although the
circuit topology and input distribution are unchanged, changing the prior changes the
magnitude---and consequently the slope---of the relationship between MMC and circuit size.
Thus, even for a fixed circuit family, the mismatch contribution reflects how well the physical
process represented by the prior is matched to the distributions encountered during
computation.

The bounds of Sec.~\ref{sec:4} are worst-case bounds and therefore do not require the actual
MMC for a particular input distribution to grow linearly with circuit size. Figure~\ref{fig_10}
provides an example in which a fixed sequence of circuits exhibits a strongly
nonlinear dependence of MMC on total gate count. This illustrates that the size-dependent
bounds derived above constrain the worst-case scaling without fixing the detailed behavior of
MMC for particular circuit families and operating distributions.

The dependence of MMC on the gate prior can also be seen already at the level of an
individual gate. Figure~\ref{fig:11} shows how the mismatch cost changes as the gate prior is
varied while the operating distribution is held fixed. Likewise, Fig.~\ref{fig:12} considers a
fixed circuit containing AND and OR gates and varies the prior associated with the OR gates.
The resulting change in total MMC illustrates how heterogeneity in the priors of different
gate types can affect the thermodynamic mismatch contribution even when the circuit
topology itself is unchanged. These examples complement Theorem~\ref{th:3}, which shows
more generally that, for heterogeneous prior bases, the relevant size-type bound is a weighted
gate count rather than a function of the total number of gates alone.

A particularly simple example shows that this dependence can reverse the ordering suggested
by circuit size. The NAND function can be implemented either by a single NAND gate or by
an AND gate followed by a NOT gate. The latter circuit contains more gates and is therefore
strictly larger. Nevertheless, if the prior associated with the NAND gate is poorly matched to
the distribution encountered during computation, while the priors associated with the AND
and NOT gates are well matched to their respective operating distributions, the two-gate
AND--NOT implementation can have lower total MMC than the single NAND gate.

Thus, minimizing circuit size and minimizing mismatch cost are, in general, distinct
optimization problems. For a fixed Boolean function and a fixed basis with priors, one may
therefore ask for a circuit family that minimizes MMC rather than size or depth, or more
generally seek architectures that optimally trade off these different resources.

\section{Discussion and Future Work}\label{sec:6}

Our results provide a framework for analyzing a particular, unavoidable contribution to the thermodynamic cost of running a Boolean circuit. For a fixed physical realization with a single island and prior distribution $q_X$, the entropy production generated from an actual initial distribution $p_X$ decomposes as
\begin{equation}
    \sigma(p_X)
    =
    \MMC(p_X)
    +
    \sigma(q_X),
\end{equation}
where $\MMC(p_X)$ is the mismatch contribution and $\sigma(q_X)\ge 0$ is the residual EP. Accordingly,
\begin{equation}
    \sigma(p_X)\ge \MMC(p_X).
\end{equation}
The residual term depends on microscopic details of the physical implementation and can vary substantially across realizations of the same logical map. In particular, the residual entropy production may be much larger than the mismatch contribution and may dominate the total entropy production. MMC should therefore be understood as a lower-bound contribution to total entropy production, rather than as the total thermodynamic cost of a circuit.

This also clarifies the sense in which MMC constitutes a minimal-cost statement. For a fixed physical process, $\MMC(p_X)$ measures the excess EP incurred when the process is run on $p_X$ rather than on its least-dissipative prior $q_X$. It is not obtained by optimizing over all possible physical implementations of the logical operation. In particular,
$q_X$ is a fixed property of the process.

This distinction is important in applications, where the operating distribution is often determined externally by environmental
sources. Such statistics determine $p_{\inn}$ and are propagated by the circuit topology to the internal gates. A physical implementation may be engineered so that its priors are well matched to an anticipated input distribution, but operating the same device under a different distribution
generates a nonzero mismatch contribution.

The circuit-level formulas derived in Sec.~\ref{sec:3} make this contribution explicit. Equation~\eqref{eq_26} expresses the MMC of gate-by-gate execution in terms of local KL-divergence and mutual-information contributions, while Eq.~\eqref{eq:MMC_layer2} gives the analogous expression for layer-by-layer execution. Once the input distribution, circuit topology, execution order, and associated priors are specified, these quantities can be evaluated without specifying a particular energy landscape, bath coupling, temperature, or continuous-time protocol. Different microscopic realizations may induce different priors, and therefore different numerical values of MMC. What is implementation-independent is the form of the decomposition once those priors are fixed.

An important consequence is that local thermodynamic optimization does not in general imply thermodynamic reversibility of the full circuit. During modular execution, the subsystem being updated can be correlated with the remainder of the circuit. Updating only that subsystem may therefore destroy or create correlations and incur mismatch cost even when the residual entropy production of the local process has been minimized.

This observation is distinct from the question of logical reversibility. A logically reversible gate can still incur mismatch cost when implemented as a local update of a correlated subsystem, while logical irreversibility does not by itself determine the entropy production of a physical implementation. Fully reversible computation constitutes a different computational model, typically involving clean ancillary bits, copying into auxiliary registers, uncomputation, and eventual restoration or resetting of those registers. Accounting for these resources requires a different thermodynamic bookkeeping and is discussed in Ref.~\cite{wolpert2019stochastic}. The present work instead concerns the mismatch cost generated by modular execution of standard Boolean circuits.

At the level of circuit families, Theorem~\ref{th:1} establishes a general upper bound on worst-case mismatch cost in terms of the local conditional priors. Full support ensures that the corresponding gate-wise constants are finite for each fixed circuit, while a uniform, $n$-independent lower bound on the conditional prior probabilities yields coefficients independent of input length. Homogeneous product priors arise as a special case. Theorem~\ref{th:3} further shows that, for heterogeneous prior bases, the relevant bound takes the form of a weighted gate count,
\begin{equation}
    \sum_{b\in\basis}\#_b(n)K_b,
\end{equation}
so the mismatch contribution can depend on gate composition as well as total circuit size.

The resulting MMC complexity classes are necessarily parameterized by a basis with priors $\widetilde{\basis}$: they characterize the scaling of worst-case mismatch cost within a specified thermodynamic circuit model, rather than assigning a hardware-independent total thermodynamic cost to a Boolean function. 

Several further directions follow naturally from this framework. One is to characterize the role of fan-in and fan-out in determining MMC. Fan-out allows a gate output to be reused at multiple downstream locations and can reduce ordinary circuit size, but its effect on mismatch
cost is less obvious because logical reuse also changes the correlation structure encountered during repeated execution. Another is to understand more systematically the dependence on execution schedule. Our result that layer-by-layer execution has lower MMC than the corresponding gate-by-gate execution suggests that parallel or temporally coarse-grained
implementations may reduce this lower-bound contribution. Whether they also reduce total entropy production in a concrete device depends on how the changed execution protocol affects its priors and residual dissipation.

A complementary direction is to instantiate the framework in explicit physical models with specified energies, reservoirs, temperatures, and continuous-time dynamics. Such a model would determine the gate priors and residual entropy productions rather than treating them as part of the circuit specification. This would permit a direct quantitative comparison between MMC and total entropy production and would reveal how tight the mismatch lower bound is for particular physical technologies.

Finally, a natural optimization problem is to identify, for a fixed Boolean function and a fixed basis with priors, circuit families that minimize mismatch cost. Comparing such families with those optimized for size or depth would clarify trade-offs among hardware resources, parallel time, and thermodynamic mismatch. Relatedly, it would be interesting to study highly parallel classes such as $\NC$ and ask whether functions admitting shallow polynomial-size circuits also admit implementations with low worst-case MMC under natural prior bases. These questions point toward a broader circuit complexity theory in which thermodynamic mismatch is treated as an additional resource rather than as a consequence
determined by size or depth alone.

\section{Acknowledgment}
This work was supported by the U.S. National Science Foundation (NSF) Grant 2221345. We thank the Santa Fe Institute for helping to support this research. We also thank Sosuke Ito, David Soloveichik and David Doty 
for helpful discussions.

\appendix
\onecolumngrid

\section{Island decomposition of the state space of a logic gate -- uniqueness and full support of the prior distributions}\label{App_prior_discussion}

A gate's state consists of the joint state of its input registers and the output register,
i.e., $g$'s state is specified by $(\x_g, \x_{\pa(g)})$. When the gate is run, the output is
changed according to the conditional distribution $\pi_g(\x_g \mid \x_{\pa(g)})$ while the
inputs remain fixed. For some state $\x'_g \in \mathrm{supp}\,\pi_g(\cdot \mid \x_{\pa(g)})$, the associated map $G$ takes the state $(\x_g, \x_\pa(g)) \to (\x'_g, \x_\pa(g))$ with probability $\pi(\x'_g | \x_{\pa(g)})$. Accordingly, the map partitions the state space into exactly $|\mathcal{X}_{\pa(g)}|$ islands, one for each parent value $\x_{\pa(g)}$. Define $\mathcal{L}(\x_{\pa(g)}) = \{(\x_g, \x_{\pa(g)}) | \x_g \in \mathcal{X}_g\}$

\[
\mathcal{X} = \bigcup_{\x_{\pa(g)} \in \mathcal{X}_{\pa(g)}} \mathcal{L}(\x_{\pa(g)}) 
\]

Note that $\mathcal{L}(\x_{\pa(g)}) \cong \mathcal{X}_g$. Let $q_{g|\x_{\pa(g)}}$ denote the prior distribution for island $\mathcal{L}(\x_{\pa(g)})$, i.e., 
\[q_{g|\x_{\pa(g)}} = \mathrm{arg}\min_{p \in \Delta_{\mathcal{L}(\x_{\pa(g)})}} \sigma(p),
\]
where $\sigma(p)$ is the EP in the execution of the gate. As discussed in Sec.~\ref{sec:2b} and proved in detail in the supplementary information of~\cite{yadav2026entropy}, the prior distribution $q_{g| \x_{\pa(g)}}$ is unique and has full support over $\mathcal{L}(\x_{\pa(g)})$. Equivalently, $q_{g|\x_{\pa(g)}}$ has full support over $\mathcal{X}_g$, i.e., 
\[
    q_{g\mid \x_{\pa(g)}}(\x_g) >0, \quad \text{for all $\x_g \in \mathcal{X}_g$}.
\]

For any initial distribution $p_{g,\pa(g)}$ over the joint state space of the gate, we can write the mismatch cost as
\begin{align}
    \MMC(p_{g,\pa(g)}) &= \sum_{\x_{\pa(g)} \in \mathcal{X}_{\pa(g)}} p_{\pa(g)}(\x_{\pa(g)})
    \left( D(p_{g\mid \x_{\pa(g)}} \| q_{g\mid \x_{\pa(g)}}) - D(\pi_{g\mid \x_{\pa(g)}} \|
    \pi_{g\mid \x_{\pa(g)}})\right) \\
    &= \sum_{\x_{\pa(g)} \in \mathcal{X}_{\pa(g)}} p_{\pa(g)}(\x_{\pa(g)})\, D(p_{g\mid
    \x_{\pa(g)}} \| q_{g\mid \x_{\pa(g)}}), \label{eq_island1}
\end{align}
where in the second KL divergence we used the observation that after execution of a gate, both
$p_{g\mid \x_{\pa(g)}}$ and $q_{g\mid \x_{\pa(g)}}$ evolve to $\pi_{g\mid \x_{\pa(g)}}$.

Moreover, define $q_{g,\pa(g)}(\x_g, \x_{\pa(g)}) = q_{g\mid\x_{\pa(g)}}(\x_g)\, q_{\pa(g)}(\x_{\pa(g)})$ for any $q_{\pa(g)}$ which has full support over $\mathcal{X}_{\pa(g)}$. Then one can re-write Eq.~\eqref{eq_island1}:
\begin{equation}\label{eq_island2}
    \MMC(p_{g,\pa(g)}) = D(p_{g,\pa(g)} \| q_{g,\pa(g)}) - D(p_{\pa(g)} \| q_{\pa(g)}) 
\end{equation}
Note that although $q_{\pa(g)}$ is required to have full support over $\mathcal{X}_{\pa(g)}$
for the expression in Eq.~\eqref{eq_island2} to be well-defined, the resulting value of
$\MMC(p_{g,\pa(g)})$ does not depend on $q_{\pa(g)}$ at all --- as is clear from
Eq.~\eqref{eq_island1}, which involves no such object. Eq.~\eqref{eq_island2} is simply a
convenient way of writing this same expression. With this in mind, in what follows we will
work with the single joint prior $q_{g,\pa(g)}$ rather than the island-wise priors
$q_{g\mid\x_{\pa(g)}}$ directly, using $q_{g,\pa(g)}$ purely as a bookkeeping device that encapsulates the (unique, full-support) island priors.

\section{Derivation of Eq.~(\ref{eq_26})}\label{app:MMC_circuits}

% \subsection{Subsystem process}\label{A1}
% Consider a system composed of two subsystems $A$ and $B$ with state spaces $\mathcal{X}_A$ and $\mathcal{X}_B$, respectively. A process that evolves an initial distribution $p^0_{AB}(x_A,x_B)$ over $\mathcal{X}_A\times\mathcal{X}_B$ is called a \textit{subsystem process} if the two subsystems evolve independently during the process,
% \begin{equation}\label{eq: sub1}
%     G_{AB}(x'_A,x'_B|x_A,x_B)=G_A(x'_A|x_A)G_B(x'_B|x_B),
% \end{equation}
% and if the entropy flow of the joint system is the sum of the entropy flows of the subsystems,
% \begin{equation}\label{eq: sub2}
%     Q(p^0_{AB})=Q_A(p^0_A)+Q_B(p^0_B).
% \end{equation}
% If $q^0_A$ and $q^0_B$ are the prior distributions for the evolution of subsystems $A$ and $B$, respectively, then the prior distribution for the joint evolution of $A$ and $B$, under conditions~\eqref{eq: sub1} and~\eqref{eq: sub2}, is the product distribution
% \begin{equation}
%     q^0_{AB}(x_A,x_B)=q^0_A(x_A)q^0_B(x_B).
% \end{equation}

Consider a circuit represented by a DAG with a set of nodes $V$ representing the Boolean gates, set of edges $E$ representing the connections between the gates, and a topological order. Let $p^t(\x)$ denote the probability that the joint state of the circuit is $\x$ before the execution of the gate $g_t$. 
Let $\widetilde{g_t}$ denote the set of all gates other than $g_t$ and $\pa(g_t)$. This partitions the set of all gates:
\begin{equation}
V = g_t \cup \pa(g_t) \cup \widetilde{g_t}.
\end{equation}
Let $p^{t+1}$ denote the probability distribution over the joint state space of the circuit after the execution of gate $g_t$. Since $g_t$ updates its state according to the conditional distribution $\pi(\x_{g_t} | \x_{\pa(g_t)})$, while the states of all other gates remain unchanged, we have
\begin{equation}\label{eq_p_t2}
    p^{t+1}(\x_{g_t},\x_{\pa(g_t)},\x_{\widetilde{g_t}}) = 
    \pi_{g_t}(\x_{g_t}|\x_{\pa(g_t)})p^t_{\pa(g_t), \widetilde{g_t}}(\x_{\pa(g_t)}, \x_{\widetilde{g_t}}),
\end{equation}

Note that since the update of gate $g_t$ depends only on the states of the gates $\pa(g_t)$,
while everything else, $\widetilde{g_t}$, remains unchanged, the update of $g_t$ constitutes
what is called a \emph{subsystem process}~\cite{wolpert2019stochastic}, with $(g_t,\pa(g_t))$
as the \emph{subsystem}. In such a subsystem process, the prior distribution factors into a
distribution over the subsystem and a distribution over the rest of the
system~\cite{wolpert2019stochastic, yadav2025minimal},
\begin{equation}\label{eq_q_t1}
    q(\x_{g_t}, \x_{\pa(g_t)}, \x_{\widetilde{g_t}}) = q_{g_t,\pa(g_t)}(\x_{g_t}, \x_{\pa(g_t)})\,
    q_{\widetilde{g_t}}(\x_{\widetilde{g_t}}),
\end{equation}
where $q_{g_t,\pa(g_t)}(\x_{g_t}, \x_{\pa(g_t)})$ is the prior distribution of the gate $g_t$ as discussed in App.~\ref{App_prior_discussion}. Once gate $g_t$ executes, this distribution evolves to
\begin{equation}\label{eq_q_t2}
    q'(\x_{g_t}, \x_{\pa(g_t)}, \x_{\widetilde{g_t}}) = \pi_{g_t}(\x_{g_t} \mid \x_{\pa(g_t)})\,
    q_{\pa(g_t)}(\x_{\pa(g_t)}) q_{\widetilde{g_t}}(\x_{\widetilde{g_t}}).
\end{equation}
The mismatch cost of step $t$ is
\begin{equation}\label{eq_B5}
    \MMC_t = D(p^t\|q) - D(p^{t+1}\|q').
\end{equation}

Using the expressions in Eqs.~\eqref{eq_p_t2}, \eqref{eq_q_t1}, and \eqref{eq_q_t2}, the KL
divergence terms can be written as
\begin{align}
    D(p^t\|q) &= I_t(X_{g_t}, X_{\pa(g_t)}; X_{\widetilde{g_t}}) +
    D(p^t_{g_t,\pa(g_t)}\|q_{g_t,\pa(g_t)}) + D(p^t_{\widetilde{g_t}}\|q_{\widetilde{g_t}}),
    \label{eq:KL0} \\
    D(p^{t+1}\|q^{t+1}) &= I_{t+1}(X_{g_t}, X_{\pa(g_t)}; X_{\widetilde{g_t}}) +
    D(p^{t+1}_{\pa(g_t)}\|q_{\pa(g_t)}) +
    D(p^{t}_{\widetilde{g_t}}\|q_{\widetilde{g_t}}), \label{eq:KL1}
\end{align}
where $I_t(X_{g_t}, X_{\pa(g_t)}; X_{\widetilde{g_t}})$ and $I_{t+1}(X_{g_t}, X_{\pa(g_t)};
X_{\widetilde{g_t}})$ denote the mutual information between the subsystem $(g_t, \pa(g_t))$
and the rest of the circuit, $\widetilde{g_t}$, before and after gate $g_t$ is executed,
respectively. Moreover, from Eqs.~\eqref{eq_actual_before} and~\eqref{eq_actual_after}, we have
\begin{align}
    p^t_{g,\pa(g)}(\x_{g_t}, \x_{\pa(g_t)}) &= p_{g_t}(\x_{g_t})\, p_{\pa(g_t)}(\x_{\pa(g_t)}) \\
    p^{t+1}_{g,\pa(g)}(\x_{g_t}, \x_{\pa(g_t)}) &= \pi_{g_t}(\x_{g_t} \mid \x_{\pa(g_t)})\,
    p_{\pa(g_t)}(\x_{\pa(g_t)}).
\end{align}
Combining these with Eqs.~\eqref{eq:KL0} and~\eqref{eq:KL1}, the mismatch cost of executing the gate can be written as
\begin{equation}\label{eq_A10}
    \MMC_t
    =
    I_t(X_{g_t},X_{\pa(g_t)};X_{\widetilde{g_t}})
    -I_{t+1}(X_{g_t},X_{\pa(g_t)};X_{\widetilde{g_t}})
    +D(p_{g_t}p_{\pa(g_t)}||q_{g_t, \pa(g_t)})
        - D(p_{\pa(g_t)}||q_{\pa(g_t)}).
\end{equation}

We further simplify the change in the mutual-information $\Delta I_t := I_t(X_{g_t},X_{\pa(g_t)};X_{\widetilde{g_t}}) - I_{t+1}(X_{g_t},X_{\pa(g_t)};X_{\widetilde{g_t}})$ using the details of dynamics of circuit implementation described in Sec.~\ref{sec3a}.

First note that we can write the mutual information terms as follows:
\begin{align}
    I_t(X_{g_t},X_{\pa(g_t)};X_{\widetilde{g_t}})&=S_{t}(X_{g_t},X_{\pa(g_t)})+S_t(X_{\widetilde{g_t}})-S_t(X), \label{eq_A14} \\
    I_{t+1}(X_{g_t},X_{\pa(g_t)};X_{\widetilde{g_t}})
    & = S_{t+1}(X_{g_t},X_{\pa(g_t)})+S_{t+1}(X_{\widetilde{g_t}})-S_{t+1}(X), \label{eq_A15}
\end{align}
where $S_t(\cdot)$ and $S_{t+1}(\cdot)$ denote the entropy of the random variable before and after the execution of gate $g_t$, respectively. 

Since before the execution, gate $g_t$ has its states from the previous run of the circuit while $\pa(g_t)$  has updated their states in the current run on new inputs, the random variable $X_{g_t}$ is independent of  $X_{\pa(g_t)}$, i.e., $S_{t}(X_{g_t},X_{\pa(g_t)}) = S_{t}(X_{g_t}) + S_{t}(X_{\pa(g_t)})$. Moreover, since the marginal distribution of the rest of the circuit $\widetilde{g_t}$ does not change during step $t$, we have $S_{t+1}(X_{\widetilde{g_t}})=S_t(X_{\widetilde{g_t}})$. Same goes for marginal distributions over $X_{g_t}$ and $X_{\pa(g_t)}$, giving us $S_t(X_{g_t}) = S_{t+1}(X_{g_t})$ and $ S_t(X_{\pa(g_t)}) = S_{t+1}(X_{\pa(g_t)})$. Using these observations, combined with the identity $S_t(X_{g_t})+S_t(X_{\pa(g_t)})-S_{t+1}(X_{g_t},X_{\pa(g_t)})
=I_{t+1}(X_{g_t};X_{\pa(g_t)}),$ we get
\begin{equation}\label{eq_A16}
    \Delta I_t
    =
    I_{t+1}(X_{g_t};X_{\pa(g_t)})+S_{t+1}(X)-S_t(X),
\end{equation}
Thus, the mismatch cost: 
    \begin{equation}
        \MMC_t
        =
        S_{t+1}(\xx) - S_t(\xx)
        + \I_{t+1}(\xx_{g_t}; \xx_{\pa(g_t)})
        + D(p_{g_t}p_{\pa(g_t)}||q_{g_t, \pa(g_t)})
        - D(p_{\pa(g_t)}||q_{\pa(g_t)}).
    \label{eq_24app}
    \end{equation}

When the $\Delta I_t$ contributions are summed over all non-input gate updates, the entropy terms telescope:
\begin{align}\label{eq_A18}
    \sum_{t=1}^{T}\Delta I_t
    &=
    S_{T+1}(X)-S_1(X)
    +\sum_{t=1}^{T}I_{t+1}(X_{g_t};X_{\pa(g_t)}).
\end{align}
Before any non-input gates are executed, the non-input nodes are remnants of the previous run, while the input nodes are freshly sampled from the new input distribution. These two collections of variables are independent at that point. After the circuit has finished updating, the non-input nodes generally become correlated with the inputs through the computation. Hence,
\begin{equation}
    S_1(X)-S_{T+1}(X)=I_0(X_{\nin};X_{\inn}).
\end{equation}
Therefore, the aggregate mismatch cost of running all non-input gates sequentially is
\begin{align}
    \sum_{t=1}^{T}\MMC_t
    &=
    -I_0(X_{\nin};X_{\inn})
    +\sum_{t=1}^{T}I_{t+1}(X_{g_t};X_{\pa(g_t)})\nonumber\\
    &\quad+
    \sum_{t=1}^{T}
    \left[
    D(p_{g_t}p_{\pa(g_t)}\|q_{g_t,\pa(g_t)})
    -D(p_{\pa(g_t)}\|q_{\pa(g_t)})
    \right].
\end{align}
The mismatch cost of overwriting the input nodes with new values sampled from $p_{\inn}$ is Eq.~\eqref{eq:overwriting_mmc}:
\begin{equation}
    \MMC_{ow}(p_\inn)=I_0(X_\inn;X_\nin)+D(p_\inn\|q_\inn).
\end{equation}
Adding the overwriting mismatch cost to the aggregate gate-update mismatch cost gives
\begin{align}
    \MMC(C_n,p_{\inn})
    &=\MMC_{ow}(p_{\inn})+\sum_{t=1}^{T}\MMC_t\\
    &=D(p_\inn\|q_\inn)
    +\sum_{t=1}^{T}I_{t+1}(X_{g_t};X_{\pa(g_t)}) +
    \sum_{t=1}^{T}
    \left[
    D(p_{g_t}p_{\pa(g_t)}\|q_{g_t,\pa(g_t)})
    -D(p_{\pa(g_t)}\|q_{\pa(g_t)})
    \right],\label{eq_A22}
\end{align}
which is Eq.~\eqref{eq_26}.

\section{Proofs}

\subsection{Proof of Thm.~\ref{th:1}}\label{proof2}
%{\color{blue}%
% We prove the base inequality~\eqref{eq:general_mmc_bound}, the linear bound~\eqref{eq:general_mmc_bound_linear}, and the product-prior special cases. To begin, we recall why the constants appearing in the bound are strictly positive.

% \paragraph*{Strict positivity of the constants (full support).}
% The prior associated with any subsystem process is the initial distribution that minimizes the associated entropy-production cost functional, which for a map $P$ and a state-dependent cost $f$ has the form $C(p)=S(Pp)-S(p)+\langle f\rangle_p$. Such a minimizer necessarily has full support within each island of the state space: for each island $c$, the minimizer $q$ satisfies $\mathrm{supp}\,q=\{x\in c:f(x)<\infty\}$, because the entropic contribution $-S(p)$ has an infinite inward directional derivative at the boundary of the simplex, ruling out any boundary minimizer. A complete and rigorous proof is given in Ref.~\cite{yadav2024mismatch} (App.~B, ``Proof that prior is always in the interior of the simplex''); this makes rigorous a property already invoked in earlier work~\cite{wolpert2020thermodynamics}. Consequently, taking the minima $q^{\min}_{\inn}$ and $q^{\min}_{g\mid\pa(g)}$ over the support of the corresponding priors (equivalently, over the island(s) visited by the actual dynamics), we have $q^{\min}_{\inn}>0$ and $q^{\min}_{g\mid\pa(g)}>0$ without any additional assumption. This is why full support is stated as a property rather than as a hypothesis of Theorem~\ref{th:1}.
% }% end blue (full-support argument)
From Eq.~\eqref{eq_26}, we have
\begin{equation}
    \MMC(C_n, p_{\inn})
        = D(p_\inn||q_\inn)
        + \sum_{t=1}^{T} \left[\I_{t+1}(\xx_{g_t}; \xx_{\pa(g_t)})
        + D(p_{g_t}p_{\pa(g_t)}||q_{g_t, \pa(g_t)})
        - D(p_{\pa(g_t)}||q_{\pa(g_t)})
        \right].
\end{equation}
Write the terms in the sum as
\begin{equation}\label{eq_26_copy}
    T_t := \I_{t+1}(X_{g_t}; X_{\pa(g_t)}) + D(p_{g_t}p_{\pa(g_t)} \| q_{g_t,\pa(g_t)}) -
    D(p_{\pa(g_t)} \| q_{\pa(g_t)}).
\end{equation}
Since mutual information is bounded above by the entropy of either marginal,
\begin{equation}
    \I_{t+1}(X_{g_t}; X_{\pa(g_t)}) \le S(X_{g_t}).
\end{equation}
Moreover, since $D(p_{g_t}p_{\pa(g_t)}|q_{g_t,\pa(g_t)}) - D(p_{\pa(g_t)}|q_{\pa(g_t)}) = \sum_{x_{\pa(g_t)}} p_{\pa(g_t)}(x_{\pa(g_t)}) D(p_{g_t} | q_{g_t|x_{\pa(g_t)}})$ from the equivalence of Eq.~\eqref{eq_island1} and Eq.~\eqref{eq_island2}, we have

\begin{align}
    T_t &\le S(X_{g_t}) + \sum_{\x_{\pa(g_t)}} p_{\pa(g_t)}(\x_{\pa(g_t)}) D(p_{g_t} | q_{g_t|\x_{\pa(g_t)}}) \nonumber \\
    &= S(X_{g_t}) - S(X_{g_t}) - \sum_{\x_{g_t},\x_{\pa(g_t)}} p_{g_t}(\x_{g_t})
    p_{\pa(g_t)}(\x_{\pa(g_t)}) \log q_{g_t\mid \x_{\pa(g_t)}}(\x_{g_t}) \nonumber \\
    &= -\sum_{\x_{g_t},\x_{\pa(g_t)}} p_{g_t}(\x_{g_t}) p_{\pa(g_t)}(\x_{\pa(g_t)}) \log
    q_{g_t\mid \x_{\pa(g_t)}}(\x_{g_t}) \nonumber \\
    &\le \log \frac{1}{q^{\min}_{g_t\mid\pa(g_t)}},
\end{align}

where the last line follows because $q_{g_t\mid\pa(g_t)}(\x_{g_t}\mid \x_{\pa(g_t)}) \ge
q^{\min}_{g_t\mid\pa(g_t)}$ for every $(\x_{g_t}, \x_{\pa(g_t)})$ in the support, so
$ -\log q_{g_t\mid\pa(g_t)}(\x_{g_t}\mid \x_{\pa(g_t)}) \le \log(1/q^{\min}_{g_t\mid\pa(g_t)})$
termwise, and the weights $p_{g_t}(x_{g_t})p_{\pa(g_t)}(\x_{\pa(g_t)})$ sum to one. Note that $q^{\min}_{g_t\mid\pa(g_t)} > 0$, so that the bound above is finite. This follows from the discussion in App.~\ref{App_prior_discussion}: for
each parent state $x_{\pa(g_t)}$, the
conditional $q_{g_t\mid\pa(g_t)}(\cdot \mid x_{\pa(g_t)})$ coincides with the prior associated
with the corresponding island, which is unique and has full support over $\mathcal{X}_{g_t}$. Therefore, the minimum of this conditional over its full support is
strictly positive for every such $x_{\pa(g_t)}$, and hence $q^{\min}_{g_t\mid\pa(g_t)} > 0$.

The input-overwriting term is bounded similarly. Since
\begin{equation}
D(p_{\inn}\|q_{\inn})
=
-S(p_{\inn})-
\sum_{x_{\inn}}p_{\inn}(x_{\inn})\log q_{\inn}(x_{\inn}),
\end{equation}
and $S(p_{\inn})\ge0$, we have
\begin{equation}
D(p_{\inn}\|q_{\inn})
\le
\log\frac{1}{q^{\min}_{\inn}}.
\end{equation}
Substituting these bounds into Eq.~\eqref{eq_26_copy}, we obtain, for any input distribution $p_{\inn}$,
\begin{equation}
\MMC(C_n,p_{\inn})
\le
\log\frac{1}{q^{\min}_{\inn}}
+
\sum_{t=1}^{T}
\log\frac{1}{q^{\min}_{g_t\mid\pa(g_t)}},
\end{equation}
which is Eq.~\eqref{eq:general_mmc_bound}. The linear bound Eq.~\eqref{eq:general_mmc_bound_linear} follows immediately from the assumed uniform bounds on the input and gate constants.

\paragraph*{Special case: product local priors.}
If the local prior of a gate $g$ factors as
\begin{equation}\label{eq:product_local_prior_special_case}
q_{g,\pa(g)}(x_g,x_{\pa(g)})
=
q_g(x_g)q_{\pa(g)}(x_{\pa(g)}),
\end{equation}
then
\begin{equation}
q_{g\mid\pa(g)}(x_g\mid x_{\pa(g)})=q_g(x_g),
\qquad
q^{\min}_{g\mid\pa(g)}=q^{\min}_g.
\end{equation}
Thus Eq.~\eqref{eq:general_mmc_bound} becomes
\begin{equation}\label{eq:product_prior_bound_general}
\MMC(C_n,p_{\inn})
\le
\log\frac{1}{q^{\min}_{\inn}}
+
\sum_{g\in V_{\nin}}
\log\frac{1}{q^{\min}_g}.
\end{equation}
Additionally, if we assume that all non-input gates share the same marginal prior, i.e., $q_g = q_\star$ for all $g \in V_{\nin}$ for some fixed $q_\star$ and the input prior factors over input bits as
\begin{equation}
q_{\inn}(x_{\inn,1},\ldots,x_{\inn,n})
=
\prod_{k=1}^n q_i(x_{\inn,k}),
\end{equation}
then
\begin{equation}
q^{\min}_{\inn}=(q_i^{\min})^n,
\qquad
q^{\min}_g=q_\star^{\min}\quad \text{for all } g\in V_{\nin}.
\end{equation}
Therefore
\begin{equation}\label{eq:MMC_C_n}
\MMC(C_n)
\le
|C_n|\log\frac{1}{q_\star^{\min}}
+n\log\frac{1}{q_i^{\min}}.
\end{equation}
In this product-prior special case the gate constant depends only on the marginal prior of the gate output, not on the number or distribution of its parents. Therefore, this homogeneous product-prior bound~\eqref{eq:MMC_C_n} is independent of the fan-in and fan-out of the gate.

% \subsection{Proof of Thm.~\ref{th:3}}\label{proof3}
% {\color{blue}%
% As noted in the main text, Theorem~\ref{th:3} follows directly from the base inequality~\eqref{eq:general_mmc_bound} of Theorem~\ref{th:1}. Under the type-dependence hypothesis, every occurrence of a gate of type $b$ carries the same conditional prior $q_{b\mid\pa(b)}$, so its contribution to the sum in Eq.~\eqref{eq:general_mmc_bound} is at most $K_b=\log(1/q^{\min}_{b\mid\pa(b)})$, with $q^{\min}_{b\mid\pa(b)}>0$ by full support (minimum taken over the support of the local prior). Summing over the $\#_b(n)$ gates of each type and retaining the input term $\log(1/q^{\min}_{\inn})$ yields Eq.~\eqref{eq:th:3}. If the input prior factorizes over the $n$ input bits with common per-bit minimum $q_i^{\min}$, then $q^{\min}_{\inn}=(q_i^{\min})^n$, so $\log(1/q^{\min}_{\inn})=nK_{\inn}$ with $K_{\inn}=\log(1/q_i^{\min})$, giving Eq.~\eqref{eq:th3_linear}. Neither the strict positivity of the $K_b$ nor the form of the input term is an independent assumption: the former is a consequence of full support, and the latter is a consequence of the factorized input prior. \hfill$\square$
% }% end blue (proof of Theorem 3)

\subsection{Circuit as a periodic process}\label{app:circuit_periodic}

A circuit can be viewed as a periodic process when a counter is integrated into its description. The counter can differentiate one clock cycle from another and sequentially run different layers or gates with each clock cycle. The periodic process, governed by the clock cycle, repeats over and over again, but sequentially implements different gates or layers with each clock cycle.

Let $Z$ denote the random variable representing the state of the counter, taking values $z \in \{0, 1, \ldots, V_{\nin} \}$. The joint state of the system is now given by $(\x, z)$, where $\x$ is the state of the circuit and $z$ is the state of the counter. We let $t$ index the sequence of gates or layers in the circuit, and $g_t$ denote the gate or layer updated at step $t$. Let $p_{XZ}(\x, z)$ denote the probability distribution over the joint state, which can be expressed as a conditional distribution of $\x$ given the state of the counter:
\begin{equation}
    p_{X, Z}(\x, z) = p_{X|Z}(\x|z)p_Z(z)
\end{equation}
At the beginning of a circuit run, the counter is initialized to $t = 0$. With each gate execution, the counter increments by one. Therefore, prior to the execution at step $t$, the counter is in state $z=t$. The distribution over circuit states just before the execution of gate $g_t$ is then given by

\begin{equation}
    p_{X,Z=t}(\x) = p_{X|Z=t}(\x),
\end{equation}
indicating that the counter is precisely in state $t$. Note that $p_{X|Z=t}(\x)$ corresponds to $p^{t}(\x)$, consistent with the notation used throughout the paper and as summarized in Table~\ref{TABLE}. 

The prior distribution $q_{X,Z}$ represents the full process that is repeated with each clock cycle of the circuit. Accordingly, the prior over the joint state $(\x, z)$ can be expressed as
\begin{equation}
q_{X, Z}(\x, z) = q_{X|Z}(\x \mid z)\, q_{Z}(z),
\end{equation}
and evolves under the periodic dynamics of the circuit to
\begin{equation}
\qqt_{X, Z}(\x, z) = \qqt_{X|Z}(\x \mid z)\, \qqt_{Z}(z),
\end{equation}
where $q'_{X|Z}$ denotes the updated distribution over circuit states following one cycle of evolution.

The prior distribution $q^{t}$, associated with the execution of gate $g_t$, is now understood as being conditioned on the counter state. In particular, the prior distribution over circuit states just before the execution of gate $g_t$ (i.e., when the counter is in state $z = t$) is given by
\begin{equation}
q_{X,Z=t}(\x) = q_{X|Z}(\x \mid z)\, q_{Z=t}(z),
\end{equation}
where $q_{Z=t}(z)=\delta_{z,t}$ is a delta distribution, enforcing that the counter is precisely in state $t$. Equivalently, this can be written more concisely as
\begin{equation}
q_{X,Z=t}(\x) = q_{X|Z=t}(\x).
\end{equation}
Applying the MMC formula to the execution of gate $g_t$, we obtain
\begin{equation}
\MMC_t = D\left(p_{X|Z=t} \| q_{X|Z=t}\right) - D\left(p_{X|Z=t+1} \| \qqt_{X|Z=t}\right),
\end{equation}
which, under the earlier notation $p^t = p_{X|Z=t}$, $q^t = q_{X|Z=t}$, and $q'^t = q'_{X|Z=t}$, becomes
\begin{equation}
\MMC_t = D(p^t \| q^t) - D(p^{t+1} \| q'^t),
\end{equation}
which is the same as Eq.~\ref{eq_B5}; further evaluation, identical to that in App.~\ref{app:MMC_circuits}, yields Eq.~\eqref{eq_26}.

\subsection{Further analysis of the mutual-information terms}
\label{app:future_mi}

The mutual-information change appearing in the derivation of
Eq.~\eqref{eq_26} admits a useful interpretation for deterministic
circuits. Recall from Eq.~\eqref{eq_A10} that
\begin{equation}
\Delta I_t
:=
I_t\!\left(
X_{g_t},X_{\pa(g_t)};
X_{\widetilde{g_t}}
\right)
-
I_{t+1}\!\left(
X_{g_t},X_{\pa(g_t)};
X_{\widetilde{g_t}}
\right).
\label{eq:deltaI_future_1}
\end{equation}
Applying the chain rule for mutual information gives
\begin{align}
\Delta I_t
={}&
I_t\!\left(
X_{\pa(g_t)};
X_{\widetilde{g_t}}
\right)
+
I_t\!\left(
X_{g_t};
X_{\widetilde{g_t}}
\mid X_{\pa(g_t)}
\right)
\nonumber\\
&-
I_{t+1}\!\left(
X_{\pa(g_t)};
X_{\widetilde{g_t}}
\right)
-
I_{t+1}\!\left(
X_{g_t};
X_{\widetilde{g_t}}
\mid X_{\pa(g_t)}
\right).
\label{eq:deltaI_future_2}
\end{align}
Neither $X_{\pa(g_t)}$ nor $X_{\widetilde{g_t}}$ changes during the
update of gate $g_t$. Therefore their joint distribution is unchanged
between steps $t$ and $t+1$, and the first and third terms in
Eq.~\eqref{eq:deltaI_future_2} cancel. Hence
\begin{align}
\Delta I_t
={}&
I_t\!\left(
X_{g_t};
X_{\widetilde{g_t}}
\mid X_{\pa(g_t)}
\right)
\nonumber\\
&-
I_{t+1}\!\left(
X_{g_t};
X_{\widetilde{g_t}}
\mid X_{\pa(g_t)}
\right).
\label{eq:deltaI_future_3}
\end{align}

For a deterministic gate, after the update
\begin{equation}
X_{g_t}=f_{g_t}\!\left(X_{\pa(g_t)}\right),
\end{equation}
so the new value of $X_{g_t}$ is completely determined by its
parents. Consequently,
\begin{equation}
I_{t+1}\!\left(
X_{g_t};
X_{\widetilde{g_t}}
\mid X_{\pa(g_t)}
\right)=0,
\end{equation}
and therefore
\begin{equation}
\Delta I_t
=
I_t\!\left(
X_{g_t};
X_{\widetilde{g_t}}
\mid X_{\pa(g_t)}
\right).
\label{eq:deltaI_future_4}
\end{equation}

To make the meaning of this term more explicit, partition
$X_{\widetilde{g_t}}$ according to the topological execution order.
Let $X_{<t}$ denote the joint state of the input nodes and non-parent
gates that have already been updated before step $t$, and let
$X_{>t}$ denote the joint state of the gates scheduled to be updated
after $g_t$. Thus,
\begin{equation}
X_{\widetilde{g_t}}
=
\left(X_{<t},X_{>t}\right).
\end{equation}
Immediately before $g_t$ is updated, $X_{<t}$ and
$X_{\pa(g_t)}$ contain values generated during the current run,
whereas $X_{g_t}$ and $X_{>t}$ still contain values inherited from
the previous run. Since the input of the current run is sampled
independently of the previous run, these two collections are
independent. In particular,
\begin{equation}
I_t\!\left(
X_{g_t};
X_{<t}
\mid X_{\pa(g_t)}
\right)=0.
\end{equation}
Using the chain rule in Eq.~\eqref{eq:deltaI_future_4} therefore gives
\begin{align}
\Delta I_t
&=
I_t\!\left(
X_{g_t};
X_{<t},X_{>t}
\mid X_{\pa(g_t)}
\right)
\nonumber\\
&=
I_t\!\left(
X_{g_t};
X_{>t}
\mid X_{\pa(g_t)},X_{<t}
\right).
\label{eq:deltaI_future_5}
\end{align}

Moreover, in the deterministic repeated-run model,
$(X_{g_t},X_{>t})$ is jointly independent of
$(X_{\pa(g_t)},X_{<t})$, since the former variables retain values
from the previous run while the latter are generated from the fresh
input of the current run. Thus the conditioning in
Eq.~\eqref{eq:deltaI_future_5} can be dropped, yielding
\begin{equation}
\boxed{
\Delta I_t
=
I_t\!\left(
X_{g_t};X_{>t}
\right).
}
\label{eq:deltaI_future_6}
\end{equation}

Equation~\eqref{eq:deltaI_future_6} gives a direct interpretation of
the information-theoretic contribution associated with updating
gate $g_t$: it is the correlation, immediately before the update,
between the gate's value inherited from the previous run and the
values of the gates that have not yet been updated. Executing
$g_t$ replaces that inherited value with one determined by its
current-run parents, thereby removing these correlations with the
remaining previous-run state.

\begin{table*}
\begin{center}
\begin{tabular}{ |p{4cm}||p{13cm}|}
 \hline
 \multicolumn{2}{|c|}{Table of notation} \\
 \hline
 \centering Symbol & Definition \\
 \hline
  \centering $C_n$& Circuit with input size $n$ \\
  \centering $g$& Index of a gate or node in the circuit, $g\in V$ \\
  \centering $t$& Execution-step index in a gate-by-gate run, $t\in\{1,\ldots,T\}$ \\
  \centering $g_t$& Gate updated at execution step $t$ \\
  \centering $\pa(g)$& Set of parent gates of gate $g$\\
  \centering $\ch(g)$& Set of children gates of gate $g$\\
  \centering $V$& Set of input nodes and non-input gates in circuit\\
  \centering $V_l$& Set of gates in layer $l$ of circuit\\
  \centering $\xx_g$& Random variable representing the state of gate $g$\\
  \centering $\xx_l$& Random variable representing the joint state of gates in layer $l$\\
  \centering $\x$& Joint state of the entire circuit\\
  \centering $\x_g$& Output state of gate $g$\\
  \centering $\x_{\pa(g)}$& Input state of gate $g$ (also, the joint state of $g$'s parents)\\
  \centering $\x_{\inn}$& Joint state of the inputs\\
  \centering $\x_{\nin}$& Joint state of non-input gates\\
  \centering $\x_{\widetilde{g_t}}$& Joint state of all gates except $g_t$ and the parents of $g_t$\\
  \centering $\xx_{<t}$& Joint state of input nodes and non-parent gates already updated before step $t$\\
  \centering $\xx_{>t}$&Joint state of gates scheduled to be updated after gate $g_t$\\
  \centering $p^t(\x)$& Distribution over circuit states before execution step $t$\\
  \centering $\qqt^t(\x)$& Distribution obtained by evolving $q^t(\x)$ through the update of gate $g_t$\\
  \centering $q_{g, \pa(g)}(\x_g, \x_{\pa(g)})$& Local prior distribution associated with gate $g$\\
  \centering $\qqt_{g, \pa(g)}(\x_g, \x_{\pa(g)})$& Distribution obtained by evolving $q_{g, \pa(g)}$ under the update of gate $g$\\
  \centering $q_{\pa(g)}(\x_{\pa(g)})$& Marginal distribution obtained from $q_{g, \pa(g)}(\x_g, \x_{\pa(g)})$\\
  \centering $\basis$& Basis of logic-gate types\\
  \centering $b$& Index of a gate type in the basis, $b\in\basis$\\
  \centering $q_b$& Shorthand for the prior distribution associated with gate type $b\in\basis$\\
  \centering $q_b^{\min}$& Probability of least likely state under distribution $q_b$\\
  \centering $\I_{t+1}(\xx_{g_t}; \xx_{\pa(g_t)})$& Mutual information between gate $g_t$ and its parents after step $t$\\
  \centering $\MMC_t$& Mismatch cost of step $t$, i.e., of updating gate $g_t$\\
  \centering $\MMC_l$& Mismatch cost of running layer $l$ in a circuit\\
  \centering $\MMC_{ow}(p_{\inn})$& Mismatch cost of overwriting inputs with input distribution $p_{\inn}$\\
  \centering $\MMC(C_n, p_{\inn})$& Total mismatch cost of running the circuit $C_n$ for the input distribution $p_{\inn}$\\
 \hline
\end{tabular}
\end{center}
\caption{Table of notation for the main symbols used in the paper.} \label{TABLE}
\end{table*}

\end{document}